\documentclass[nofootinbib,prd,aps,superscriptaddress,preprintnumbers]{revtex4}
\pdfoutput=1
\usepackage{amsmath,amssymb,euscript}
\usepackage{color}
\usepackage{accents}
\usepackage{hyperref}
\usepackage{ulem}
\usepackage{epsfig}
\usepackage{varioref}

\def\beq{\begin{equation}}
\def\eeq{\end{equation}}

\newcommand{\tev}{\,\, \mathrm{TeV}}
\newcommand{\gev}{\,\, \mathrm{GeV}}

\renewcommand{\emph}{\textit}

\graphicspath{{figs/}}

\begin{document}

\begin{flushright}
PITT-PACC-1507
\end{flushright}

\title{Integrating in the Higgs Portal to Fermion Dark Matter}

\def\Cincy{Department of Physics, University of Cincinnati, Cincinnati, Ohio 45221,USA}
\def\Pitt{Pittsburgh Particle physics, Astrophysics \& Cosmology Center (PITT-PACC),
Department of Physics \& Astronomy, University of Pittsburgh, Pittsburgh, PA 15260, USA}
\author{Ayres Freitas}
\email[Electronic address:]{afreitas@pitt.edu}
\affiliation{\Pitt}
\author{Susanne Westhoff}
\email[Electronic address:]{suw22@pitt.edu}
\affiliation{\Pitt}
\author{Jure Zupan}
\email[Electronic address:]{zupanje@ucmail.uc.edu}
\affiliation{\Cincy}

\vspace{1.0cm}
\begin{abstract}
\vspace{0.2cm}\noindent 
Fermion dark matter (DM) interacting with the standard model through a Higgs portal requires non-renormalizable operators, signaling the presence of new mediator states at the electroweak scale. Collider signatures that involve the mediators are a powerful tool to experimentally probe the Higgs portal interactions, providing complementary information to strong constraints set by direct DM detection searches. Indirect detection experiments are less sensitive to this scenario.
We investigate the collider reach for the mediators using three minimal renormalizable models as examples, and requiring the fermion DM to be a thermal relic. The Large Hadron Collider in its high-energy, high-luminosity phase can probe most scenarios if DM is lighter than about $200\gev$. Beyond this scale, future high-energy experiments such as an electron-positron collider or a 100-TeV proton-proton collider, combined with future direct detection experiments, are indispensable to conclusively test these models.
\end{abstract}
\maketitle


\section{Introduction}
Dark matter (DM) can couple to a gauge invariant operator $H^\dagger H$, where $H$ is the standard model (SM) Higgs field. Since $H^\dagger H$ is the lowest-dimensional Lorentz- and gauge-invariant operator in the SM, such Higgs portal couplings to DM could dominate the interactions between the visible sector and the dark sector~\cite{Patt:2006fw, MarchRussell:2008yu,Andreas:2008xy, Englert:2011yb, Lebedev:2011iq, LopezHonorez:2012kv,
  Djouadi:2012zc, Greljo:2013wja, Fedderke:2014wda, Craig:2014lda}. If DM is a scalar, $\varphi$, the Higgs portal operator, $(\varphi^\dagger \varphi)(H^\dagger H)$, is renormalizable and the theory is in principle UV-complete. In contrast, if DM is a fermion, $\chi$, the Higgs portal interactions are of mass dimension five,
\beq
{\cal L}_{\rm EFT}=\frac{\kappa_1}{\Lambda}(\bar \chi \chi) (H^\dagger H)+\frac{i \kappa_5}{\Lambda}(\bar \chi \gamma_5 \chi) (H^\dagger H).
\label{eq:eft}
\eeq
This necessarily implies the existence of new states at an energy scale $\Lambda$, the mediators between the Higgs field and the DM. Integrating out the mediators gives the above interactions in an Effective Field Theory (EFT). The presence of mediators offers new opportunities to experimentally search for the Higgs portal: Instead of searching directly for DM, one can search for the mediators instead. 

The mediators in general fall into two categories: {\it  i)} mediators that are electroweak singlets, or {\it ii)} mediators that are charged under the electroweak gauge group. In this paper we explore the phenomenology of both categories using three minimal renormalizable models of the Higgs portal (henceforth called UV completions). As a representative of the first category we choose a model where both the DM and the mediator are electroweak singlets ({\it the singlet-singlet model}). For the second category we consider two examples, a model where DM can be an electroweak singlet ({\it the singlet-doublet model}), and a model where DM is necessarily part of an electroweak multiplet ({\it the doublet-triplet model}).

For each model we compute the predicted relic density from thermal freeze-out and determine the parameter space that is consistent with the observed dark matter relic density. The viable parameter regions are confronted with bounds from direct detection, indirect detection and collider experiments. Interestingly, we find that the combination of these constraints often requires the mediators to have masses of similar magnitude as the DM fermion, $\chi$. This means that the EFT Lagrangian in Eq.~\eqref{eq:eft} is in general not a good description for the computation of the relic density and the collider phenomenology. For these observables, we therefore treat all fields in the dark sector as dynamical degrees of freedom, i.e. we ``integrate in'' the Higgs portal.

Finally, we analyze how the UV Higgs portal completions can be probed by future direct detection and collider experiments. In particular, we consider the upcoming 14-TeV run of the Large Hadron Collider (LHC) with up to 3000~fb$^{-1}$ luminosity, a future $e^+e^-$ collider with a center-of-mass energy of up to $\sqrt{s}=1\tev$, and a future $pp$ collider with $\sqrt{s}=100\tev$.

The paper is organized as follows. In Sec.~\ref{sec:models} we introduce the three minimal UV completions of the fermion DM Higgs portal, in Sec.~\ref{sec:relic-density} we calculate the respective thermal relic densities, and in Sec.~\ref{sec:direct-detection} we estimate the DM direct detection rates. Sec.~\ref{sec:gamgam} deals with the constraints following from Higgs decays. In Sec.~\ref{sec:coll} we combine all the above constraints with the expected sensitivity of the LHC 14-TeV run and future colliders on the mediators. Appendix~\ref{app:one-loop-dd} contains the analytic formulae for loop-induced DM-Higgs interactions in a version of the doublet-triplet model that give the dominant direct detection signal.


\section{The models}
\label{sec:models}

We start by introducing three minimal UV completions of the Higgs portal models with fermionic DM: (A)  the singlet-singlet model, (B) the singlet-doublet model, and (C) the doublet-triplet model. All three examples are treated as \emph{simplified models} rather than full theories, i.e.\ we do not consider issues such as anomaly cancellation and stability under renormalization group running to high scales.

\subsection{The Singlet-Singlet Model}
The dark sector is assumed to consist of DM, which is a $\mathbb{Z}_2$-odd SM-singlet Majorana fermion, $\chi$, and of a mediator, which is a $\mathbb{Z}_2$-even real singlet scalar, $\phi_S$. All SM fields are $\mathbb{Z}_2$ even. This model has already been discussed at length in the literature, see for example Refs.~\cite{Kim:2008pp,Baek:2011aa,Baek:2012uj,Fairbairn:2013uta,Esch:2013rta,Baek:2014jga,Bagherian:2014iia}. In this paper we update the limits on the model and show projections for the 14-TeV LHC and a future 100-TeV $pp$ collider. 

The relevant terms for interactions between the dark sector and the SM in the Lagrangian are
\beq
\begin{split}
{\cal L}_{\rm S}\supset &-\mu_0^2|H|^2 - \lambda_0|H|^4 - \tfrac{1}{2}m^2_{S,0}\phi_S^2 - V_0(\phi_S) - \mu'_0|H|^2\phi_S - \lambda'_0|H|^2\phi_S^2\\
&- \tfrac{1}{2}m_{\chi,0}\bar\chi\chi - \tfrac{1}{2}y_0\bar \chi\chi \phi_S - \tfrac{i}{2}y_{5,0}\bar \chi\gamma_5\chi \phi_S,
\end{split}
\eeq
where $H$ is the SM Higgs field and $V_0(\phi_S)$ contains cubic and quartic $\phi_S$ self-interactions. Here we use the four-component notation for the Majorana fermion $\chi$. The scalar fields $H$ and $\phi_S$ acquire the vacuum expectation values (vevs) $v=246$~GeV and $v_S$, respectively. The vev $v$ breaks the electroweak symmetry, while $v_S$ does not. We can therefore write
\beq
H = (G^+,(v+h+iG^0)/\sqrt{2})^\top, \qquad \phi_S = v_S + S,
\eeq
with $G^+, G^0$ the would-be Goldstone bosons eaten by the longitudinal components of the $W^+$ and $Z$ bosons, respectively. 
In terms of $h$ and $S$ the interaction Lagrangian is
\beq
\begin{split}\label{eq:ss-lag}
{\cal L}_{\rm S}\supset & \ \mu^2\Big(h^2 + \frac{h^3}{v} + \frac{h^4}{4v^2}\Big) - \tfrac{1}{2}m^2_SS^2 - V(S) - 
(\mu' S + \lambda' S^2)\bigl (vh + \tfrac{1}{2}h^2 \bigr )\\
&- \tfrac{1}{2}m_{\chi}\bar\chi\chi - \tfrac{1}{2}y\bar\chi\chi S- \tfrac{i}{2}y_5\bar\chi\gamma_5\chi S,
\end{split}
\eeq
where $m_\chi =m_{\chi,0}+ y_0v_S$ is the DM mass,
$y=y_0$ and $y_5=y_{5,0}$ are the parity-conserving and parity-violating Yukawa couplings, respectively,
  and $m_S^2 = m^2_{S,0} + \lambda'_0v^2 + V''_0(v_S)$ is the singlet mass squared. The Higgs--singlet mixing parameter is given by $\mu' = \mu'_0 + 2\lambda'_0v_S$, while $\mu^2 = \mu^2_0 + \mu'_0v_S + \lambda'_0v_S^2,$ and $\lambda^{(')}= \lambda_0^{(')}$. We also define 
 $V(S)=\mu_S S^3+\lambda_S S^4$ that contains triple and quartic singlet scalar interactions.

The DM state $\chi$ interacts with the SM through Higgs-singlet mixing. For $\mu'\ne0$, the mass eigenstates $h_{1,2}$ are admixtures of $h$ and $S$, 
\beq\label{eq:mixingSh}
h_1=c_\alpha h-s_\alpha S, \qquad h_2=s_\alpha h+ c_\alpha S.
\eeq
Here we use the abbreviations $c_\alpha= \cos\alpha$, $s_\alpha= \sin\alpha$. The mixing angle $\alpha$ and the masses of $h_{1,2}$ are given by
\beq\label{eq:mixingSh2}
s_\alpha^2 = \frac{1}{2} \biggl(1 - \frac{m_S^2+2\mu^2}{\Delta m^2} \biggr), \qquad m^2_{h_{1,2}} = \frac{1}{2} \Big(m_S^2-2\mu^2 \mp \Delta m^2 \Big),
\eeq
with the mass splitting $(\Delta m^2)^2=(m_S^2+2\mu^2)^2 + 4\mu'^2v^2$. The state $h_1$ is the observed SM-like scalar with mass $m_{h_1}= 125.09\pm 0.24\gev$~\cite{Aad:2015zhl}, whereas $h_2$ is mostly singlet-like. The mixing angle $\alpha$ is constrained by the measured Higgs production and decay rates and, depending on the mass $m_{h_2}$, by the non-observation of a second Higgs scalar at the LHC, as we show in Sec.~\ref{sec:coll-ss}.
In particular, Higgs--singlet mixing leads to the DM--SM interactions 
(here and below $\hat{\cal L}$ denotes Lagrangian in the mass eigenstate basis)
\beq
\mathcal{\widehat{L}} \supset -\tfrac{1}{2}y \bigl( -s_\alpha \, \bar\chi\chi h_1 + c_\alpha \, \bar\chi\chi h_2 \bigr)-
\tfrac{i}{2} y_5 \bigl( -s_\alpha \, \bar\chi\gamma_5\chi h_1
 + c_\alpha \, \bar\chi\gamma_5\chi h_2 \bigr)
   - \sum_f \frac{y_f}{\sqrt{2}} \bigl ( c_\alpha \, \bar ff h_1 + s_\alpha \, \bar ff h_2 \bigr),
\eeq
where $y_f$ is the SM Yukawa coupling of the fermion $f$. The scalar interactions also govern direct detection signatures of spin-independent DM scattering off nuclei. Since the relevant effective coupling $\chi\chi \bar q q$ is $\sim \sin(2\alpha)$ (see Tab.~\ref{tab:dd-couplings}), direct detection experiments set strong constraints on the mixing angle $\alpha$ (see Figs.~\ref{fig:ssh} and \ref{fig:singlet-singlet}).

\subsection{The Singlet-Doublet Model}\label{sec:sd}
In this model the dark sector consists of two fermion fields, $\chi_D$ and $\chi_S$, transforming under the SM electroweak group $SU(2)_L\times U(1)_Y$ as 
\beq\label{eq:chiD}
\chi_D\sim (2, 1/2),\qquad \chi_S\sim (1,0).
\eeq
The field $\chi_D=(\chi_D^+,\chi_D^0)$ is a doublet of Dirac fermions with vector-like gauge interactions, while the singlet $\chi_S$ can be either a Dirac or a Majorana fermion. We discuss both of these possibilities.  In the dark sector we impose a $\mathbb{Z}_2$ symmetry, $\chi_{D,S}\to -\chi_{D,S}$, under which the SM fermions are even. This forbids mixing with neutrinos and makes the lighter of the two mass eigenstates in the dark sector stable. 

\vspace{1em} 
\paragraph{Dirac singlet fermion.} 
We first discuss the case where $\chi_S$ is a Dirac fermion. The particle content of this model consists of two neutral Dirac fermions, $\chi_S$ and $\chi_D^0$, and a charged Dirac fermion $\chi_D^+$. The relevant terms in the Lagrangian are 
\beq\label{eq:sd-yukawa}
{\cal L}_{\rm m}\supset -m_D\bar \chi_D\chi_D-m_S\bar \chi_S\chi_S - \big(y \bar \chi_D \chi_S H+\text{h.c.}\big).
\eeq
Without loss of generality, the Yukawa coupling $y$ can be chosen real by redefining the complex phase of the $\chi_D$.
After electroweak symmetry breaking (EWSB) this term introduces the mixing between $\chi_S$ and $\chi_D^0$.
For later convenience, we define the mixing angle $\theta_a$ generally as
\beq\label{eq:mixing-angle}
\sin^2\theta_a = \frac{1}{2} \biggl(1 + \frac{m_D-m_S}{\Delta m_a}\biggr),\qquad\text{with}\quad (\Delta m_a)^2 = (m_S-m_D)^2 + a(yv)^2\,.
\eeq
Here $a=2$, and the heavy and light mass eigenstates, $\chi_h^0$ and $\chi_l^0$, are given by
\beq\label{eq:sD:mixing}
\chi_h^0=\cos\theta_2\, \chi_S+\sin\theta_2\, \chi_D^0, \qquad \chi_l^0=-\sin\theta_2\, \chi_S+\cos\theta_2\, \chi_D^0,
\eeq
with the corresponding mass eigenvalues
\beq
m_{h,l}^0 = \tfrac{1}{2}\big(m_D+m_S \pm \Delta m_2 \big).
\eeq
The charged state $\chi_D^+$ has a mass $m^+=m_D$. In the mass eigenstate basis, the interactions of the neutral fermions with the $Z$ and Higgs bosons read
\begin{align}\label{eq:sd-intmix}
\mathcal{\widehat{L}} \supset & - \frac{g}{2c_W}\Big[\cos^2\theta_2\, \bar \chi_l^{0} \gamma^\mu \chi_l^0 + \sin^2\theta_2\, \bar \chi_h^{0} \gamma^\mu \chi_h^0 + \frac{1}{2}\sin(2\theta_2)\,\big(\bar \chi_h^{0} \gamma^\mu \chi_l^0 + \bar \chi_l^{0} \gamma^\mu \chi_h^0\big)\Big]Z_\mu\\\nonumber
 & \ - \frac{y}{\sqrt{2}} \Big[ \sin(2\theta_2)\, \big(\bar \chi_h^0\chi_h^0 - \bar \chi_l^0\chi_l^0\big) + \cos(2\theta_2)\, \big(\bar \chi_h^0\chi_l^0 + \bar \chi_l^0\chi_h^0\big) \Big] h.
\end{align}
It is interesting to consider two parameter limits of this model. For $|m_D-m_S|\ll |yv|$, $\chi_S$ and $\chi_D^0$ are maximally mixed, $\theta_2\approx \pi/4$. The two neutral mass eigenstates are split by $m_{h}^0-m_{l}^0\approx \sqrt2|yv|$, while the charged state has a mass $m_+ = m_D \approx (m_h^0+m_l^0)/2$, and thus lies in between the two neutral states. Because of the large mixing, the DM coupling to the $Z$ boson is unsuppressed, so that direct DM detection searches exclude this possibility.

If $|m_D - m_S| \gg yv$, $\chi_h^0$ is significantly heavier than $\chi_l^0$. For $m_D>m_S$, the DM state is mostly a singlet, $\chi_l^0\simeq -\chi_S+\theta' \chi_D^0$ with $\theta'\simeq |yv|/\sqrt2 (m_D-m_S)$. The coupling of DM to the $Z$ boson is thus suppressed, and the model is allowed by direct DM searches.
In contrast, if $m_S>m_D$, DM is mostly a doublet, $\chi_l^0\sim \chi_D^0$, with unsuppressed couplings to the $Z$ boson. This possibility is therefore excluded by direct DM searches. 

\vspace{1em} 
\paragraph{Majorana singlet fermion.} \label{majsd}
The second possibility is that $\chi_S$ is a Majorana fermion. This scenario
corresponds to the bino-higgsino system (with decoupled wino) in the Minimal Supersymmetric Standard Model (MSSM) for $\tan\beta=1$ and $y=g'/\sqrt{2}$. Some phenomenology of the singlet-doublet Majorana model has also been explored in Refs.~\cite{Carena:2004ha,Mahbubani:2005pt,D'Eramo:2007ga,Enberg:2007rp,Cohen:2011ec}\footnote{During the final stages of our work, a phenomenological study of the Majorana singlet-doublet model with an analysis of collider constraints became available in Ref.~\cite{Calibbi:2015nha}.}. We will slightly abuse the notation, such that in this subsection $\chi_S$ denotes a two-component Weyl fermion, while $\chi_{D}$ and $\chi_{D}^{c}$ are two left-handed Weyl fermions forming a Dirac fermion. They transform under $SU(2)_L\times U(1)_Y$ as $\chi_D\sim (2, 1/2)$ and $\chi_D^c \sim (2, -1/2)$. The translation to four-component notation is given by $(\chi_{D}, \epsilon \chi_{D}^{c\ast})\to \chi_D$, where the final $\chi_D$ is the Dirac fermion in Eq.~\eqref{eq:chiD} (here $\epsilon_{ij}$ is the antisymmetric tensor in the $SU(2)_L$ space).
In two-component notation, the relevant terms in the Lagrangian read
\beq\label{eq:Majorana:singlet}
{\cal L}_m\supset m_D\chi_{D}^c\epsilon\chi_{D} -\tfrac{1}{2}m_S\chi_S\chi_S - y (H^\dagger \chi_{D} \chi_S  - \chi_S  \chi_{D}^c \epsilon H) + \text{h.c.},
\eeq
where the contractions of Lorentz and $SU(2)_L$ indices are implicit (for Lorentz contractions we use the convention of Ref.~\cite{Dreiner:2008tw}). 
Here and henceforth we assume that $\chi_D$ and $\chi_D^c$
 couple with equal but opposite strength to the Higgs boson,\footnote{One could write the Yukawa interaction more generally as 
 $y_L H^\dagger \chi_{D} \chi_S  - y_R \chi_S  \chi_{D}^c \epsilon H $. By choosing the phase of $\chi_S$ in Eq.~\eqref{eq:Majorana:singlet}, $m_S$ can be made real and positive. Furthermore, by adjusting the phases of $\chi_{D}$ and $\chi_{D}^c$, we can make $m_D$ and one of the Yukawa couplings, $y_L$ or $y_R$, positive and real. Here we assume that $y_L=y_R=y$, a choice motivated by reducing the corrections to the $T$ parameter. Notice that changing the sign of $m_D$ is equivalent to changing the sign of $\chi_D$ (or $\chi_{D}^c$), which implies changing the sign of $y_L$ (or $y_R$).
This results in replacing $-m_D$ with $m_D$ and replacing $yv/\sqrt2$ with
 $- yv/\sqrt2$ (or replacing $-yv/\sqrt2$ with
 $yv/\sqrt2$) in Eq.~\eqref{eq:M0:doublet-singlet}.}
 and choose $y$ to be real.
This prevents contributions to the electromagnetic dipole moment of the electron~\cite{Mahbubani:2005pt}. Furthermore, the interactions in Eq.~\eqref{eq:Majorana:singlet} feature a global $SU(2)_R$ symmetry (broken by the hypercharge), which protects the electroweak $T$ parameter from large corrections.
 After EWSB, the mass term for the neutral states is given by ${\cal L}_m=-\frac{1}{2}({\cal M}_{0})_{ij} \chi_i \chi_j +{\rm h.c.}$.
 In the basis $\chi_i=\{\chi_S, \chi_{D}^{c 0 }, \chi_{D}^0\}$, the mass matrix reads
\beq\label{eq:M0:doublet-singlet}
{\cal M}_0 = \begin{pmatrix}m_S & -\frac{yv}{\sqrt{2}} & \frac{yv}{\sqrt{2}} \\ 
- \frac{yv}{\sqrt{2}} & 0 & -m_D \\
 \frac{yv}{\sqrt{2}} & -m_D & 0 \end{pmatrix},
\eeq
which is diagonalized by the following transformation
\beq\label{eq:smd-eigensystem}
 \begin{pmatrix}\chi_h^0\\ \chi_m^0 \\
\chi_l^0 \end{pmatrix} =\begin{pmatrix}\cos\theta_4 & -\frac{1}{\sqrt{2}}\sin\theta_4 & \frac{1}{\sqrt{2}}\sin\theta_4 \\
0 & \frac{i}{\sqrt{2}} & \frac{i}{\sqrt{2}} \\
\sin\theta_4 & \frac{1}{\sqrt{2}}\cos\theta_4 & -\frac{1}{\sqrt{2}}\cos\theta_4 \end{pmatrix} \begin{pmatrix}\chi_S\\
\chi_{D}^{c0} \\ \chi_{D}^{0} \end{pmatrix}.
\eeq
The mixing angle $\theta_4$ is given by Eq.~(\ref{eq:mixing-angle}) with $a=4$.
The masses of the three neutral eigenstates, $\chi_{h,l}^0$ and $\chi_m^0$, are
\beq
m_{h,l}^0=\tfrac{1}{2}\big(m_D+m_S\pm \Delta m_4 \big), \qquad m_m^0=m_D,
\eeq
respectively, while the mass of the charged state $\chi_D^+$ is $m_+=m_D$ as in the Dirac $\chi_S$ case. The couplings of the neutral fermions to $Z$ and $h$ are given by 
\begin{align}
{\cal \widehat{L}} \supset & \ i \frac{g}{2c_W} \big(\sin\theta_4\, \chi_h^{0\ast} - \cos\theta_4\, \chi_l^{0\ast}\big) \bar \sigma^\mu \chi_m^0 Z_\mu + {\rm h.c.}\\\nonumber
& - \frac{y}{2} \Big[\sin(2\theta_4)\big(\chi_h^{0}\chi_h^0 - \chi_l^{0}\chi_l^0\big) - 2 \cos(2\theta_4)\chi_h^{0}\chi_l^0 \Big]h + {\rm h.c.}.
\end{align}
Let us consider the parameter limits in this model. For $m_D\approx m_S$ (i.e. $|m_D-m_S|\ll 2yv$, but $m_D \gtrsim 2yv$), the mixing angle is $\theta_4\simeq \pi/4$. We have three Majorana states split by $|yv|$, that is $m_{h,l}^0\simeq m_D\pm |yv|$ and $m_m^0=m_D$. The lightest state, $\chi_l^0$, is the DM candidate. All $Z$ couplings to the DM field are thus off-diagonal.
The DM direct detection signal due to tree-level $Z$ exchange is therefore kinematically forbidden, as long as the mass splitting $\Delta m_4$ is larger than several hundred keV. This requirement is fulfilled for all the benchmarks that we consider.
At the same time, in the above limit of $\theta_4\simeq \pi/4$ the coupling of the Higgs boson to the DM field $\chi_l^0$ is maximal and dominates spin-independent DM interactions with nuclei.

In the limit $|m_D-m_S|\gg yv$ with $m_D \gg m_S$, we have $\theta_4 \simeq \pi/2$. The states $\chi_h^0$ and $ \chi_m^0$ are quasi-degenerate with masses $m_h^0=m_D+(yv)^2/m_D$ and $m_m^0=m_D$, respectively, and form a pseudo-Dirac state. Its couplings to the $Z$ boson are that of the neutral component in a Dirac fermion electroweak doublet, if the mass splitting can be ignored. The lightest state $\chi_l^0$ is mostly a singlet with mass $m_l^0=m_S- (y v)^2/m_D$. In the decoupling scenario $m_D\gg m_S,yv$, both the $Z$ and Higgs exchange in direct DM searches are thus absent at the tree level. This feature protects the decoupling scenario from being excluded by direct DM searches. If instead the singlet decouples, i.e. if $m_S\gg m_D$, then the mixing angle is $\theta_4\simeq 0$. Now $\chi_l^0$ and $\chi_m^0$ are quasi-degenerate states with mass $m_l^0\approx m_m^0=m_D$. They form a pseudo-Dirac fermion with unsuppressed couplings to the $Z$ boson. This limit is therefore excluded by direct DM searches.

For $m_D \lesssim 2yv$, the lightest neutral state can be a pure doublet, $\chi_m^0$. In this scenario, DM interactions with nuclei are absent at tree level. Up to radiative corrections the DM state $\chi_m^0$ is mass-degenerate with the charged state $\chi_D^+$, which leads to strong co-annihilation (see Sec.~\ref{sec:relic-density}).

\subsection{The Doublet-Triplet Model}\label{sec:doublet:triplet}
In this model we assume that the dark sector consists of two fermions, an electroweak doublet, $\chi_D$, and a triplet, $\chi_T$, that transform under $SU(2)_L\times U(1)_Y$ as
\beq
\chi_D\sim (2, 1/2+r),\qquad \chi_T\sim (3,r).
\eeq
We consider the cases $r=0,-1$, in which both $\chi_D$ and $\chi_T$ have neutral components. For $r=0$, $\chi_T$ can be either a Dirac fermion or a Majorana fermion, while $\chi_D$ is always a Dirac fermion. We thus consider three cases: $r=-1$ with Dirac triplet, $r=0$ with Dirac triplet, and $r=0$ with Majorana triplet.

\setcounter{paragraph}{0}
\vspace{1em} 
\paragraph{Dirac triplet fermion, $r=-1$.} In this case the electroweak triplet is composed of a neutral state, $\chi_T^0$, a singly charged state, $\chi_T^-$, and a doubly charged state, $\chi_T^{--}$, while the electroweak doublet contains a neutral state, $\chi_D^0$, and a charged state, $\chi_D^-$, thus
\beq
 \chi_D = \begin{pmatrix}\chi_D^0\\ \chi_D^- \end{pmatrix},\quad \chi_T = \begin{pmatrix}
\chi_T^-/\sqrt{2} & \chi_T^0 \\ \chi_T^{--} & -\chi_T^-/\sqrt{2}\end{pmatrix}.\label{eq:dtm1}
\eeq
  The mass terms and Yukawa interactions are 
 \beq\label{eq:mass:triplet}
{\cal L}_m\supset -m_D\bar \chi_D\chi_D - m_T\,{\rm Tr}\big(\bar \chi_T\chi_T \big) - \big(y \bar \chi_D \chi_T H+\text{h.c.}\big).
\eeq
After EWSB, the corresponding mass matrices in the bases $\{\chi_T^0, \chi_D^0\}$ and $\{\chi_T^-, \chi_D^-\}$ are 
\beq\label{eq:mass:matrix:triplet}
{\cal M}_0 = \begin{pmatrix}
 m_T & y\frac{v}{\sqrt{2}}\\
 y\frac{v}{\sqrt{2}} & m_D \end{pmatrix},\qquad
 {\cal M}_- =  \begin{pmatrix}m_T & -y\frac{v}{2}\\ -y\frac{v}{2} & m_D \end{pmatrix},
\eeq
respectively, while the mass of $\chi_T^{--}$ is $m_T$. EWSB thus introduces mixing between both neutral and charged states, so that the mass eigenstates are given by
\begin{align}
\chi_h^0 & =\cos\theta_2\, \chi_T^0+\sin\theta_2\, \chi_D^0, \qquad\quad \chi_h^+ = \cos\theta_1\, \chi_T^++\sin\theta_1\, \chi_D^+,\\\nonumber
\chi_l^0 & =-\sin\theta_2\, \chi_T^0+\cos\theta_2\, \chi_D^0, \qquad \chi_l^+ =-\sin\theta_1\, \chi_T^++\cos\theta_1\, \chi_D^+,
\end{align}
with the corresponding mass eigenvalues
\beq
m_{h,l}^0 = \tfrac{1}{2}\big(m_D+m_T \pm \Delta m_2 \big), \qquad m_{h,l}^+ = \tfrac{1}{2}\big(m_D+m_T \pm \Delta m_1 \big).
\eeq
The mixing angles $\theta_{2,1}$ and the mass splittings $\Delta m_{2,1}$ are defined in (\ref{eq:mixing-angle}), with $m_S\rightarrow m_T$.
Since the splitting in the neutral sector is larger than in the charged sector, $\Delta m_2 > \Delta m_1$, the lightest neutral state $\chi_l^0$ is a potential DM candidate if $m_D$ and $m_T$ have the same sign. As in Sec.~\ref{sec:sd}, we use the freedom in the phase of $\chi_D$ to make $y$ real and positive. In the basis of mass eigenstates, the interactions of the neutral states with the $Z$ and Higgs bosons are given by
\beq
\begin{split}
\mathcal{\widehat{L}} \supset &\ \frac{g}{c_W}\Big[\big(1-\frac{\sin^2\theta_2}{2}\big)\,\bar\chi_h^0\gamma^\mu\chi_h^0 + \big(1-\frac{\cos^2\theta_2}{2}\big)\,\bar\chi_l^0\gamma^\mu\chi_l^0 - \frac{\sin2\theta_2}{4}\,\big(\bar\chi_h^0\gamma^\mu\chi_l^0 + \bar\chi_l^0\gamma^\mu\chi_h^0 \big)\Big]Z_\mu\\
&-\frac{y}{\sqrt{2}} \Big[ \sin(2\theta_2)\, \big(\bar \chi_h^0\chi_h^0 - \bar \chi_l^0\chi_l^0\big) + \cos(2\theta_2)\, \big(\bar \chi_h^0\chi_l^0 + \bar \chi_l^0\chi_h^0\big) \Big]h.
\end{split}
\eeq
We consider two parameter limits, {\it i)} almost degenerate doublet and triplet, and {\it ii)} the decoupling limit.
The degenerate case occurs for $m_D\approx m_T$, so that $|m_D-m_T|\ll |yv|$. In this case the mixing is maximal in both the neutral and charged sectors, $\theta_{2,1}\simeq \pi/4$. The mass eigenvalues are $m_{h,l}^0 \simeq m_T \pm {|yv|}/{\sqrt{2}}$ for the neutral states, and $m_{h,l}^- \simeq m_T \pm {|yv|}/{2}$ for the charged states. 

In the decoupling limit $m_D,m_T\gg yv$, DM is either predominantly the neutral component of the doublet (for $m_D<m_T$) or of the triplet (for $m_T<m_D$), with a mixing angle of ${\mathcal O}(y^2 v^2/(m_T-m_D)^2)$.  The masses for the neutral states in the two cases are $m_h^0 \simeq m_{D(T)} + {y^2v^2}/{(2|m_D-m_T|)}$ and $m_l^0 \simeq m_{T(D)} - {y^2v^2}/{(2|m_D-m_T|)}$, while for the charged states the mass deviation from $m_{D,T}$ is half as large. Since in this model the coupling of DM to the $Z$ boson is not suppressed, the direct DM detection searches exclude the entire region of parameter space that is consistent with the observed DM relic density.

\vspace{1em} 
\paragraph{Dirac triplet fermion, $r=0$.} The electroweak doublet is composed of a neutral state, $\chi_D^0$, and a charged state, $\chi_D^+$, while the electroweak triplet is composed of one neutral state, $\chi_T^0$, and two charged states, $\chi_T^+, \chi_T'^-$,
\beq
 \chi_D = \begin{pmatrix}\chi_D^+\\ \chi_D^0 \end{pmatrix},\quad \chi_T = \begin{pmatrix}
\chi_T^0/\sqrt{2} & \chi_T^+ \\ \chi_T'^{-} & -\chi_T^0/\sqrt{2}\end{pmatrix}.\label{eq:dt0}
 \eeq
 The mass terms and Yukawa couplings are given in Eq.~\eqref{eq:mass:triplet}. After EWSB, the mass matrices for neutral and charged states are similar to Eq.~\eqref{eq:mass:matrix:triplet}, but with ${\cal M}_0\to {\cal M}_+$ and ${\cal M}_-\to {\cal M}_0$, in the bases $\{ \chi_T^+, \chi_D^+\}$ and $\{\chi_T^0, \chi_D^0\}$, respectively, while $\chi_T'^-$ has a mass of $m_T$.
The heavy and light states in the neutral and charged sectors have masses
\beq
m_{h,l}^0 = \tfrac{1}{2}\big(m_D+m_T \pm\Delta m_1 \big), \qquad m_{h,l}^+ = \tfrac{1}{2}\big(m_D+m_T \pm\Delta m_2 \big).
\eeq
The lightest neutral state can thus be lighter than the lightest charged state, if $m_D$ and $m_T$ have opposite signs. However, as for the case $r=-1$, this model is excluded by constraints from the DM relic density and direct detection searches. In particular, obtaining the correct thermal relic density requires sizeable doublet-triplet mixing (see Sec.~\ref{sec:relic-density}),  which leads to large direct detection rates from coupling of the doublet component to the $Z$ boson.

\vspace{1em} 
\paragraph{Majorana triplet fermion, $r=0$.}
Taking instead the electroweak triplet to be a Majorana fermion, i.e.\ $\chi_T^\pm = \chi_T'^\pm$ in (\ref{eq:dt0}), while the doublet is still a Dirac fermion, the mass terms and Yukawa interactions in the two-component notation are (see also Eq.~\eqref{eq:Majorana:singlet})
\beq
{\cal L}_m\supset m_D\chi_{D}^c\epsilon\chi_{D} -\frac{1}{2}m_T \text{Tr}\big(\chi_T \chi_T\big) - y (H^\dagger \chi_{T} \chi_D  -  \chi_{D}^{c\top} \epsilon \chi_T  H) + \text{h.c.},
\eeq
where, as in the Majorana singlet-doublet model, we have introduced two left-handed Weyl fermions transforming under $SU(2)_L\times U(1)_Y$ as $\chi_D\sim (2, 1/2)$ and $\chi_D^c \sim (2, -1/2)$. As for the Majorana singlet-doublet model, the form of the Yukawa coupling terms is restricted by a global $SU(2)_R$ symmetry to protect the electroweak $T$ parameter. We use the freedom of field redefinitions to make $m_D$ and $y$ real and positive.\footnote{The sign of $y$ can be flipped by adjusting the phases of $\chi_D$ and $\chi_D^c$, without affecting other terms in the Lagrangian. Physical observables are thus insensitive to $y\leftrightarrow -y$.}
The mass $m_T$ is in general a free complex parameter, but restricted to real values in our analysis.
After EWSB, the Yukawa interactions mix the triplet and the doublet components. The three neutral Weyl fermions $\chi_i^0=\{\chi_T^0, \chi_{D}^{c0},\chi_{D}^0\}$ have a Majorana mass term $\mathcal{L}_{\rm m}\supset-\frac12 ({\cal M}_0)_{ij}\chi_{i}^{0}\chi_j^0+{\rm h.c.}$. 
 The charged fermions have Dirac masses $\mathcal{L}_m\supset-\chi_{m}^- ({\cal M}_+)_{mn}\chi_n^++ \text{h.c.}$,
 where the negatively charged Weyl fermions are $\chi_m^-=\{\chi_T^-,\chi_{D}^{c-}\}$, and the positively charged ones, $\chi_n^+=\{\chi_T^+, \chi_{D}^+\}$.
The two mass matrices are
\beq
{\cal M}_0 = \begin{pmatrix}m_T & y\frac{v}{2} & -y\frac{v}{2} \\ 
y\frac{v}{2} & 0 & -m_D \\
-y\frac{v}{2} & -m_D & 0 \end{pmatrix},
\qquad
{\cal M}_+ = \begin{pmatrix}
m_T & y\frac{v}{\sqrt{2}}\\ 
y\frac{v}{\sqrt{2}} & m_D 
\end{pmatrix}.
\eeq
The mass matrix for neutral states is diagonalized by, cf. Eq.~\eqref{eq:smd-eigensystem},
\beq\label{eq:tmd-eigensystem}
 \begin{pmatrix}
 \chi_{a}^0\\ 
 \chi_{ b}^0 \\
\chi_{c}^0 
\end{pmatrix} =
\begin{pmatrix}\cos\theta_2 & \frac{1}{\sqrt{2}}\sin\theta_2 & -\frac{1}{\sqrt{2}}\sin\theta_2 \\
0 & \frac{i}{\sqrt{2}} & \frac{i}{\sqrt{2}} \\
\sin\theta_2 & -\frac{1}{\sqrt{2}}\cos\theta_2 & \frac{1}{\sqrt{2}}\cos\theta_2 
\end{pmatrix} 
\begin{pmatrix}
\chi_T^0\\
\chi_{D}^{c 0}\\ \chi_{D}^{0}
\end{pmatrix},
\eeq
whereas the charged mass eigenstates are
\beq
\chi^+_a = \cos\theta_2\chi_T^+ + \sin\theta_2\chi_D^+, \qquad \chi_c^+ = -\sin\theta_2\chi_T^+ + \cos\theta_2\chi_D^+\,.
\eeq
The mass spectrum is given by
\beq\label{eq:dmt-mass}
m^0_{a,c} = m^+_{a,c}=\tfrac{1}{2}\big(m_D+m_T \pm \Delta m_2 \big),
\qquad
m^0_{b} = m_D.                                                              
\eeq
In the basis of mass eigenstates, the couplings of neutral fermions to the $Z$ and Higgs bosons are given by
\begin{align}\label{eq:maj-dt-int}
{\cal {\widehat{L}}} \supset &  \ -i \frac{g}{2c_W} \big(\sin\theta_2\, \chi_{a}^{0\ast} - \cos\theta_2 \,\chi_{ c}^{0\ast}\big) \bar \sigma^\mu \chi_{ b}^0 Z_\mu + {\rm h.c.}\\\nonumber
 & -\frac{y}{2\sqrt{2}} \Big[ \sin(2\theta_2)\,\big(\chi_{ a}^{0}\chi_{ a}^0 - \chi_{ c}^{0}\chi_{ c}^0\big) - 2\cos(2\theta_2)\, \chi_{a}^{0}\chi_{c}^0  \Big]h + {\rm h.c.}.
\end{align}
This scenario corresponds to the wino-higgsino system (with decoupled bino) in the MSSM for $\tan\beta=1$ and $y=g$. The DM phenomenology of the doublet-triplet Majorana model for the case where $0 < m_D \lesssim 200\gev$ and $y \gtrsim 1$ has been studied in Ref.~\cite{Dedes:2014hga}.\footnote{Our
notation corresponds to the one used in Ref.~\cite{Dedes:2014hga} for $\chi_1^0 = -i\chi_b^0$, $\chi_2^0=\chi_c^0$, $\chi_3^0=\chi_a^0$, $\chi_1^+=\chi_c^+$, $\chi_2^+=-\chi_a^+$, and $M_D = -m_D$.} In this
parameter region, one typically has the mass ordering $m_b < |m_{a,c}|$, so that the DM candidate $\chi^0_l = \chi^0_b$ is a pure doublet fermion. In this case $\chi^0_l$ has no diagonal tree-level couplings to the $Z$ and Higgs bosons. The direct detection cross-section in this scenario is loop-induced and thus suppressed (see Sec.~\ref{sec:direct-detection}).

On the other hand, if $m_D,\,m_T \gg 200\gev$, the mass ordering is $m_c < m_b < m_a$ and the lightest neutral and charged states are mass-degenerate at tree-level. This degeneracy is lifted by one-loop corrections involving gauge bosons, leading to~\cite{Giudice:1995np,Cheng:1998hc,Feng:1999fu,Gherghetta:1999sw}
\beq\label{eq:rad:split}
{\cal M}_+ = \begin{pmatrix}m_T+\delta m_T^+ & y\frac{v}{\sqrt{2}}\\ y\frac{v}{\sqrt{2}} & m_D+\delta m_D^+ \end{pmatrix}.
\eeq
In the limit $m_T,m_D,|m_T-m_D| \gg m_Z$ the radiative splittings are given by
\beq
\delta m_T^+ = \frac{g^2}{8\pi}(m_W-c_W^2m_Z), \qquad
\delta m_D^+ = \frac{e^2}{8\pi}m_Z.
\eeq
The corrections to the off-diagonal elements in the mass matrix have been neglected above, which is justified for $yv \ll |m_T-m_D|$. The one-loop corrections can also be neglected for the calculation of the mixing angles. Since the corrections $\delta m_T^+$ and $\delta m_D^+$ are positive, the lightest state of the spectrum is the neutral state $\chi_c^0$, a DM candidate. As is apparent in Eq.~\eqref{eq:maj-dt-int}, the $Z$ boson coupling to a DM pair $\chi_c^0\chi_c^0$ is absent and the Higgs coupling is proportional to $\sin(2\theta_2)$. Direct detection signals are thus suppressed for a small mixing angle $\theta_2$.


\section{Thermal relic density}\label{sec:relic-density}
As described above, we assume that the fermionic Higgs portal is responsible for explaining the entire dark matter density through thermal freeze-out in the early universe.

In the singlet-singlet model, DM preferentially annihilates
into $h_2h_2$ and $h_2h_1$ final states, if kinematically allowed. The amplitude for $\chi\chi \to h_2h_2$ is proportional to $\cos^2\alpha$, where $\alpha$ is the $h$--$S$ mixing angle, so that annihilation can be efficient even for very small values of $\alpha$. For $m_\chi < (m_{h_1}+m_{h_2})/2$, the main annihilation channels are into $h_1h_1$ and $W^+W^-$, where the latter proceeds 
via off-shell $h_{1,2}$ exchange in the $s$-channel. For small values of $m_\chi$, the $b\bar{b}$ final state can also become relevant. The rates for these processes grow with $\sin\alpha$. The requirement of a sufficiently large annihilation cross-section then imposes a lower bound on $\sin\alpha$.

\begin{figure}
\hfill
\includegraphics[height=0.7in]{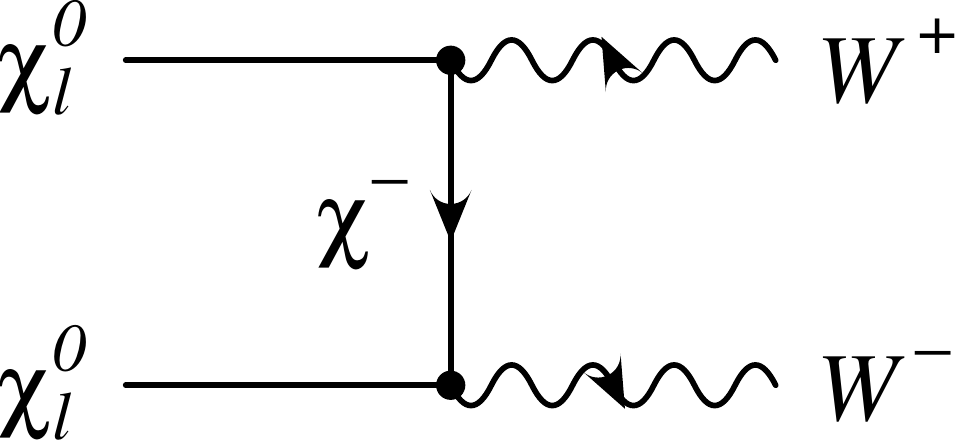}
\hfill
\includegraphics[height=0.7in]{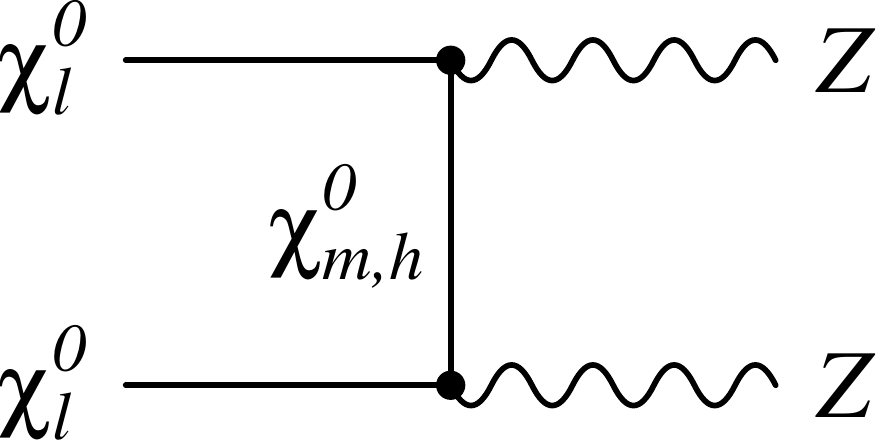}
\hfill
\includegraphics[height=0.7in]{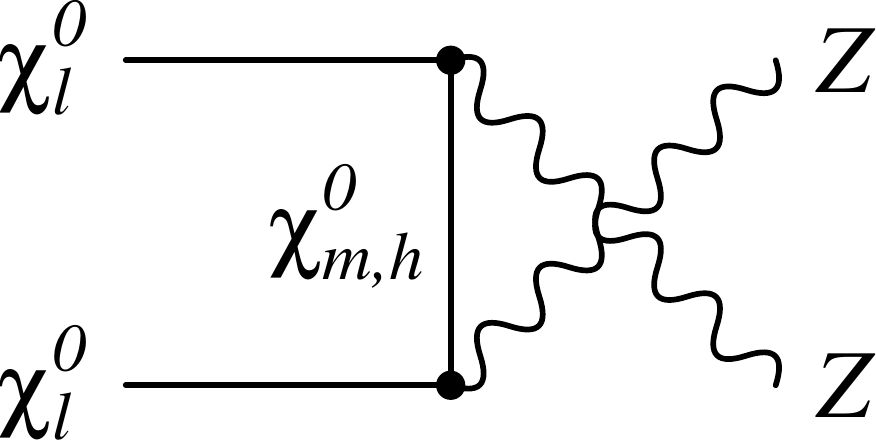}
\hfill {}
\vspace{-1ex}
\caption{Feynman diagrams for the dominant annihilation channels in the Majorana DM models with fermion mediators.}
\label{fig:diag1}
\end{figure}

For the Majorana DM models with fermion mediators (the Majorana singlet-doublet model in Sec.~\ref{sec:sd} and the Majorana doublet-triplet model in Sec.~\ref{sec:doublet:triplet}),
the main channels for pair annihilation of the neutral DM candidates involve $WW$ and $ZZ$ final states. These processes are mediated by one of the
mediator fermion states in the $t$-channel, see Fig.~\ref{fig:diag1},  since the $Z\chi^0_l\chi_l^0$ coupling vanishes exactly. 
For the Majorana singlet-doublet model, annihilation via the $s$-channel Higgs-boson resonance is also a viable option for $m_l^0 \approx m_h/2$. In this case, resonant enhancement from the Higgs-boson propagator leads to a sufficiently large annihilation cross-section to produce the correct relic density. The dominant annihilation final states are then given by the leading Higgs decay modes, $i.\,e.$ $b\bar{b}$, $W W^*$, $gg$ and $\tau^+\tau^-$.

In contrast, in the Dirac singlet-doublet model, the DM annihilation mainly proceeds through $s$-channel $Z$-boson exchange. Only for very large DM masses, $m^0_l \sim {\cal O}$(1 TeV), annihilation into $WW$ and $ZZ$ final states through $t$-channel fermion exchange becomes important.

In the singlet-doublet models (both for the Majorana and Dirac cases), the lightest neutral fermion is constrained to be mostly singlet, to avoid the strong direct detection bounds for doublet dark matter (see Sec.~\ref{sec:direct-detection}). However, the singlet nature of DM in these models also suppresses the annihilation cross-section, thus typically yielding too large of a relic density. Nevertheless, the correct DM density could still be obtained if 
$\chi^0_l\chi^\pm$ and $\chi^0_l\chi^0_m$ co-annihilation 
processes contribute at a sizeable level. As a result, the allowed parameter space is limited to relatively small values for the mass splitting $m_D-m_S$.
One exception is the Higgs resonance region for the Majorana singlet-doublet model, where the correct value for the annihilation cross-section can be obtained without co-annihilation.

In the scenario with pure doublet Majorana DM, $\chi_m^0$, the degeneracy of the states $\chi_m^0$ and $\chi_D^+$ leads to strong co-annihilation. The correct thermal relic density is obtained only if all dark particles lie above the TeV scale. Since such a scenario is not accessible at the LHC or planned future colliders, e.g., a proton-proton collider with $\sqrt{s}\leq 100\tev$ or an $e^+e^-$ collider with $\sqrt{s}\leq 1\tev$, we do not investigate this possibility further.

We have computed the relic density using {\tt MicrOMEGAs 3.6.9.2}~\cite{Belanger:2013oya}, which automatically incorporates co-annihilation processes and three-body final states through off-shell $W/Z$ production.
The models described in the previous section have been implemented into model files for {\tt CalcHEP}~\cite{Belyaev:2012qa}, which is used for the matrix-element generation within {\tt MicrOMEGAs}. The relic density is then required to match the value obtained by the Planck collaboration~\cite{Planck:2015xua},
\beq
\Omega_{\rm DM} h^2 = 0.1199 \pm 0.0022.
\label{eq:cdm}
\eeq
For the models with fermion mediators, this reduces the three-dimensional parameter space, e.g.\ $m_D,\,m_{S/T},\,y$, for each model down to two independent parameters. This is illustrated by the solid colored lines in Figs.~\ref{fig:majsd}--\ref{fig:majdt}.

It is worth pointing out that the Majorana doublet-triplet model has two distinct regions of parameter space that are compatible with the relic density constraint from Eq.~\eqref{eq:cdm}. The first region is realized for $m_D \lesssim 200\gev$ and $y \gtrsim 1$, leading to a DM candidate $\chi^0_l=\chi_b^0$ that is an unmixed doublet state with mass of a few 100~GeV~\cite{Dedes:2014hga}.

The second possibility is obtained for $m_D, m_T \gtrsim 1\tev$. In this case, the lightest neutral and charged states, $\chi^0_l = \chi^0_c$ and $\chi_l^\pm = \chi^\pm_c$, are split only by small radiative corrections (see the end of Sec.~\ref{sec:doublet:triplet}). As a result, there is strong co-annihilation and the relic density comes out too small if the DM mass is below about 1~TeV. 

\begin{figure}
\includegraphics[height=.45\textwidth]{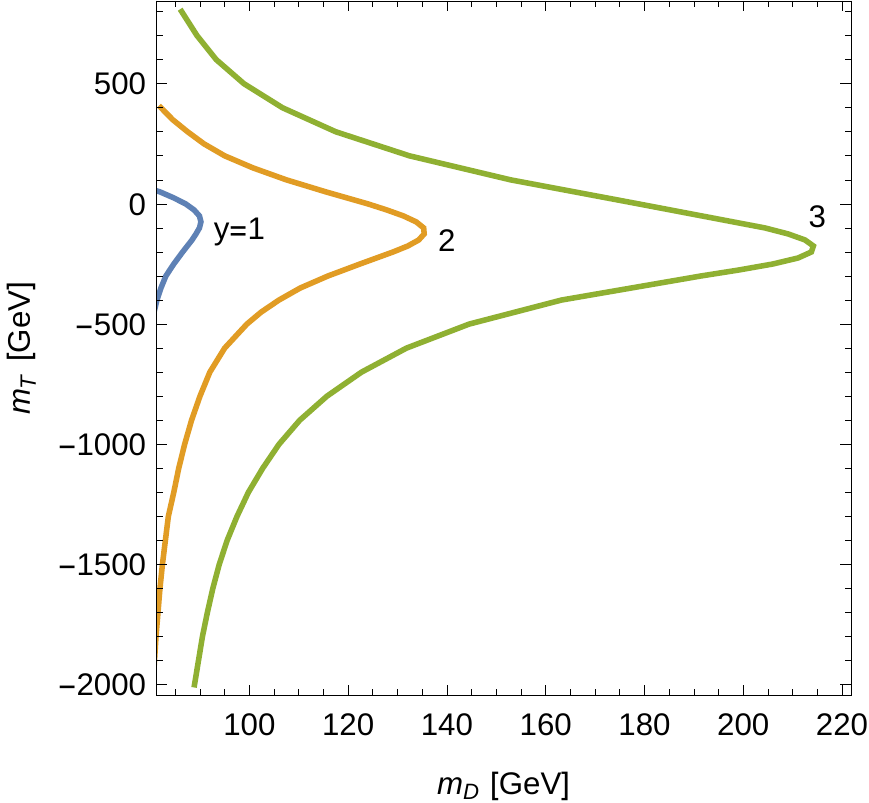}
\includegraphics[height=.45\textwidth]{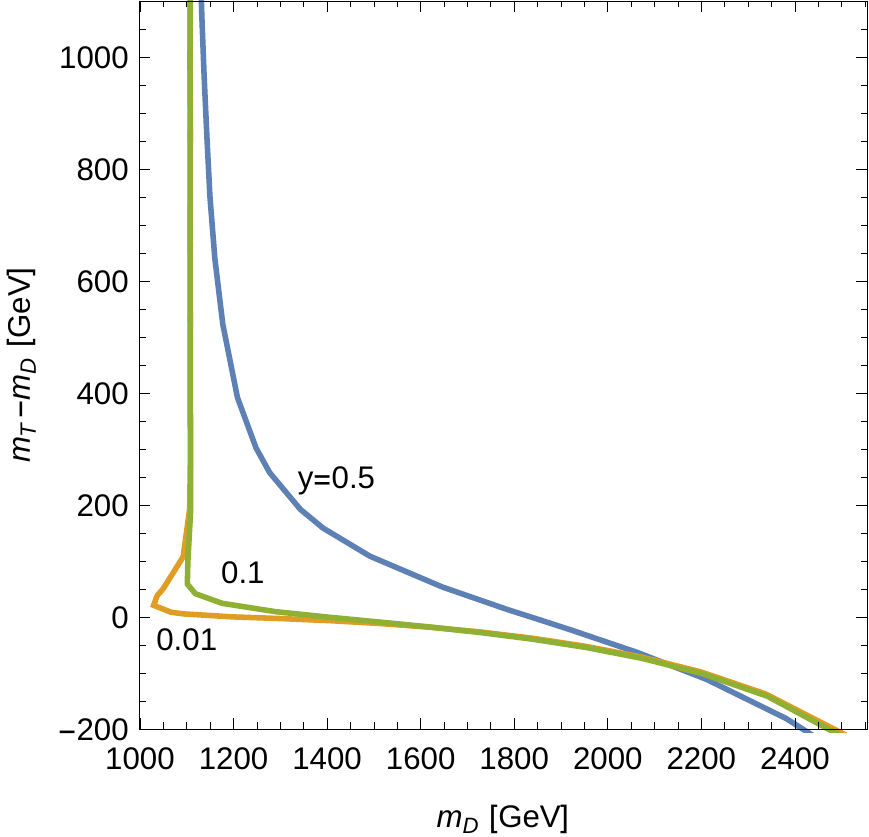}~ 
\vspace{-1ex}
\caption{Range of Lagrangian parameters $m_D$ and $m_T$ compatible with the thermal relic density constraints from Eq.~\eqref{eq:cdm} in the Majorana doublet-triplet model. The curves correspond to different values of the Yukawa coupling $y$.}
\label{fig:majdtx}
\end{figure}

Fig.~\ref{fig:majdtx} shows the parameter space that is compatible with the relic density constraint in both cases, as a function of the mass parameters $m_{D,T}$ and for several sample values of $y$. To ensure perturbativity, an upper limit $y < 3$ has been imposed.
Unfortunately, the second scenario, where all the new fermion states are beyond 1~TeV, is inaccessible at the LHC and planned future accelerators. Therefore we will not study it any further in Sec.~\ref{sec:coll}.


\section{Direct detection}\label{sec:direct-detection}
The non-observation of DM scattering off atomic nuclei, referred to as direct detection, leads to strong bounds on our models of fermion DM. At the tree level, spin-independent interactions of the lightest neutral state of the spectrum, $\chi_l^0$, with quarks $q$ inside the nucleus are generally mediated by the Higgs boson and, in case $\chi_l^0$ is a Dirac fermion, also by the $Z$ boson. Since the momentum transfer in DM-nucleus scattering is much smaller than the Higgs or $Z$ mass, spin-independent DM-quark interactions can be described by an effective Lagrangian,
\beq\label{eq:dd-eff}
\widehat{\mathcal{L}}_{\rm EFT}^{\chi q} \supset - G_Z^q\,(\bar{\chi}_l^0\gamma_{\mu}\chi_l^0)\,(\bar{q}\gamma^{\mu}q) - G_h^q\,(\bar{\chi}_l^0\chi_l^0)\,(\bar{q}q)\,.
\eeq
The $Z$- and Higgs-mediated effective couplings $G_Z^q$ and $G_h^q$ are listed for our models in Tab.~\ref{tab:dd-couplings}. We show only models with a suppressed $Z$ coupling, which can pass the limits from direct detection searches and simultaneously provide the correct DM relic density. By comparing the dependence on the fermion mixing angle $\theta_a$ with the definition in (\ref{eq:mixing-angle}), it is apparent that all models favor a small Yukawa coupling $y$ to evade bounds from direct detection.
\begin{table}[tb]
\centering
\begin{tabular}{ccccc}
\hline\hline
Model & \,DM state\, & $G_Z^q$ & $G_h^q$ & $f_n/f_p\approx$ \\
\hline
\,Majorana $\chi_S + S$ & $\chi_S$ & -- & $\frac{g m_q}{8m_W}y\sin(2\alpha)\bigl(m_{h_1}^{-2}-m_{h_2}^{-2}\bigr)$ & 1 \\
\hline
 Dirac $\chi_S + \chi_D$ & $\chi_S$ & $-\frac{g^2(T_q^3-2s_W^2Q_q)}{4c_W^2m_Z^2}\cos^2\theta_2$~~&~~$\,\frac{g}{2m_h^2}\frac{m_q}{m_W}\frac{y}{\sqrt{2}}\sin(2\theta_2)$~~&~~$-\frac{1}{1-4s_w^2}$ \\
\,Majorana $\chi_S + \chi_D\,$ & $\chi_S$ & -- & $\frac{g}{4 m_h^2}\frac{m_q}{m_W}y\sin(2\theta_4)$ & $1$\\
\hline
\,Majorana $\chi_T + \chi_D\,$ & $\chi_D^0$ & -- & $0 +\frac{g}{4 m_h^2}\frac{m_q}{m_W}\delta y$ & $1$\\
\,Majorana $\chi_T + \chi_D\,$ & $\chi_T^0$ & -- & $\,\frac{g}{4 m_h^2}\frac{m_q}{m_W}\frac{y}{\sqrt{2}}\sin(2\theta_2)$ & $1$\\
\hline\hline
\end{tabular}
\caption{Effective DM-quark interactions from $Z$ exchange ($G_Z^q$) and Higgs exchange ($G_h^q$). The column ``DM state'' indicates the dominant component of the lightest neutral state in each model. The ratio $f_n/f_p$ gives the amount of isospin violation, assuming that $Z$ exchange dominates. Furthermore, $m_h$ is the Higgs mass, $g$ is the weak coupling constant, and $T_q^3$ and $Q_q$ are the weak isospin three-component and the electric charge of the quark $q$. The one-loop correction $\delta y$ is given in Appendix~\ref{app:one-loop-dd}.}
\label{tab:dd-couplings}
\end{table}
 At zero momentum transfer, the cross-section for spin-independent scattering of a DM particle off a nucleus with mass number $A$ and proton number $Z$ is given by
\beq
\sigma_A = \frac{k\mu_A^2}{\pi}\big\vert Z f_p+(A-Z) f_n\big\vert ^2\ \stackrel{f_p=f_n}{\longrightarrow}\ \sigma_{p}\frac{\mu_A^2}{\mu_p^2}A^2,
\eeq
where $f_p$ and $f_n$ are the DM couplings to protons and neutrons, $\sigma_{p}=k\mu_p^2f_p^2/\pi$ is the DM-proton cross-section, $\mu_i^2= m_\chi^2 m_i^2/(m_\chi+m_i)^2$ is the reduced mass, and $k=1(4)$ if DM is a Dirac (Majorana) fermion. In terms of the effective interactions in Eq.~(\ref{eq:dd-eff}), the DM-proton coupling is given by
\beq
f_p = \delta_{D} \sum_{q} G_Z^q + \sum_{q} f_{Tq}^{(p)} G_h^q \frac{m_{p}}{m_q} + \frac{2}{27}f_{TG}^{(p)}\sum_{Q} G_h^Q\frac{m_{p}}{m_Q},
\eeq
and analogously for $f_n$ with $p\to n$. Here $\delta_{D}=1\,(0)$ for a Dirac (Majorana) DM fermion, $q=u,d,s$ and $Q=c,b,t$ denote light and heavy quarks, and $f_{TG}^{(p)} = 1-\sum_{u,d,s} f_{Tq}^{(p)}$, where $m_pf_{Tq}^{(p)} = \langle p|m_q\,\bar q q|p\rangle$ describes the matrix element of quarks inside the proton.

The results of direct detection searches are usually expressed in terms of the DM-nucleon cross-section $\sigma_{N}=k\mu_N^2f_N^2/\pi$, taking the isotope abundances in the detector material into account and assuming isospin-conserving couplings $f_N=f_p=f_n$ and $\mu_N^2=\mu_p^2\approx\mu_n^2$. However, if DM is a Dirac fermion, the interaction through $Z$ exchange can induce significant isospin violation, resulting in $f_p\neq f_n$. When deriving bounds on the DM-proton cross-section $\sigma_p$ in our Dirac DM models, we therefore rescale the experimental bounds on $\sigma_N$ from Ref.~\cite{Feng:2011vu}
\beq\label{eq:iso-viol}
\frac{\sigma_{p}}{\sigma_{N}}=\frac{\sum_i\eta_i\mu_{A_i}^2A_i^2}{\sum_i\eta_i\mu_{A_i}^2|Z+(A_i-Z)f_n/f_p|^2},
\eeq
where $\eta_i$ denotes the abundance of isotope $A_i$. The ratio $f_n/f_p$ for our models is listed in the last column on Tab.~\ref{tab:dd-couplings}.
The $Z$-exchange contribution in the Dirac singlet-doublet model is so large that it dominates the DM-quark interaction. The resulting isospin breaking leads to a cancellation in the DM-nucleus scattering amplitude. The bounds from direct detection on this model are therefore much weaker than they would be when neglecting isospin effects. Since isospin violation in Higgs exchange is very small, the cross-sections in all other models are not subject to isospin breaking effects.

In our numerical analysis, we compare the DM-nucleon cross-sections computed with {\tt MicrOMEGAs 3.6.9.2}~\cite{Belanger:2013oya} with the latest bounds from the LUX experiment~\cite{Akerib:2013tjd} and projected bounds for the XENON1T experiment~\cite{Aprile:2012zx}. We use the values for the scalar nucleon quark form factors $f_{Tq}^{(p,n)}$ given in Tab.~3 of Ref.~\cite{Belanger:2013oya}\footnote{It should be kept in mind, however, that the values for $f_{Ts}^{(p,n)}$ are subject to large theoretical uncertainties~\cite{Young:2013nn,Junnarkar:2013ac}. Hadronic uncertainties on $f_{Tu}^{(p,n)}$ and $f_{Td}^{(p,n)}$ can be reduced by using the method described in Ref.~\cite{Crivellin:2013ipa}.}. For the Dirac singlet-doublet model we take into account isospin violation using (\ref{eq:iso-viol}), with the isotope abundances for Xenon listed in Tab. II of Ref.~\cite{Feng:2011vu}. The resulting bounds on the parameter space in each model are displayed in the figures in Sec.~\ref{sec:coll}.

In models with Majorana DM, $Z$-mediated spin-independent scattering off nuclei is generally absent at tree level, since the Majorana fermion is its own anti-particle, while the vector current is odd under charge conjugation. In the Majorana doublet-triplet model with pure doublet DM, $\chi_l^0=\chi_b^0$, also the Higgs-mediated scattering vanishes at tree level. The reason is that $\chi_b^0$ does not obtain its mass from the Higgs mechanism and does not mix with the triplet through Yukawa interactions. A DM-Higgs interaction is induced at one-loop through the exchange of dark fermions and electroweak gauge bosons, 
\begin{equation}
\widehat{\mathcal{L}} = -\frac{\delta y}{2}\,\chi_b^0\chi_b^0 h + {\rm h.c.}.
\end{equation}
The Yukawa correction $\delta y$ is a function of the parameters $y$, $m_a^0=m_a^+$, and $m_c^0=m_c^+$. An analytic expression is given in Appendix~\ref{app:one-loop-dd}.
 This vertex generates an effective DM-nucleon scalar interaction $G_h^q$ through Higgs exchange, as given in Tab.~\ref{tab:dd-couplings}. Scalar interactions are also induced by box diagrams with $W$ and $Z$ bosons. These contributions are numerically sub-leading if the Yukawa coupling $y$ is large, which is the case if DM is a doublet. We therefore neglect them in our analysis.\\


\section{Higgs decays}\label{sec:gamgam}
In the singlet-singlet model, mixing between the doublet Higgs $h$ and the singlet scalar $S$ leads to a universal suppression of all couplings of the SM-like scalar $h_1$,
\beq
\kappa_t=\kappa_b=\kappa_\tau=\kappa_W=\kappa_Z=\cos\alpha,
\eeq
where $\kappa_i \equiv g_{h_1ii}/g_{hii}^{\rm SM}$.
Thus the value of the mixing angle $\alpha$ can be constrained from measurements of the Higgs production rates at the LHC and future colliders \cite{Lopez-Val:2013yba,Englert:2014uua}.

In the doublet-triplet model, the decay rate of the Higgs boson into two photons, $\Gamma(h\to \gamma\gamma)$, is changed at the one-loop level
due to the presence of virtual charged fermions. The decay ratio with respect to the SM rate, $\Gamma_{\rm SM}(h\to\gamma\gamma)$, is given by
\beq
R_\gamma = \frac{\Gamma(h\to \gamma\gamma)}{\Gamma_{\rm SM}(h\to \gamma\gamma)}=\Big\vert 1 + \frac{A_\chi}{A_{\rm SM}}\Big\vert^2.
\eeq
Here the one-loop amplitudes for the SM, $A_{\rm SM}$, and the charged fermions, $A_{\chi}$, are defined as
\beq
A_{\rm SM} = \sum_f N_c Q_f^2 A_F(\tau_f)+A_B(\tau_W)\qquad \text{and}\qquad A_{\chi} = \sum_{\chi} Q_\chi^2 y_\chi \frac{v}{m_\chi} A_F(\tau_\chi),
\eeq
where $\tau_i=m_h^2/(4m_i^2)$ and the loop functions for $\tau\leq 1$ are,
\beq
\begin{aligned}
A_F(\tau) & = \frac{2}{\tau^2}\Big\{\tau+(\tau-1)\arcsin^2\sqrt{\tau}\Big\},\\
A_B(\tau) & = -\frac{1}{\tau^2}\Big\{2\tau^2+3\tau+3(2\tau-1)\arcsin^2\sqrt{\tau}\Big\}.
\end{aligned}
\eeq
In terms of the mass eigenstates from Eq.~(\ref{eq:dmt-mass}), the amplitude is given by
\beq
A_\chi = \frac{y\sin(2\theta_2)}{\sqrt{2}}\Big\{\frac{v}{m_a^+} A_F(\tau_{\chi^+_a}) - \frac{v}{m_c^+} A_F(\tau_{\chi^+_c})\Big\},\qquad \sin(2\theta_2) = \frac{\sqrt{2}yv}{m_a^+-m_c^+}.
\eeq
Notice that $A_{\chi}$ is proportional to $y^2$ and thus independent of the sign of the Yukawa coupling. Furthermore, the contribution vanishes in the limit of no fermion mixing, so that at least two charged fermions are required to induce the Higgs decay. Since $A_F(\tau)$ is monotonously increasing with $\tau$ and since $m_a^+ > m_c^+$ per definition, $A_{\chi}$ is always negative for positive $m_{a,c}^+$. The contribution of charged fermions thus depletes the $h\to\gamma\gamma$ decay rate below the SM expectation. As we will discuss in Sec.~\ref{sec:coll}, the measurements of $R_\gamma$ by the LHC collaborations set strong constraints on the Majorana triplet-doublet model in the parameter space where the Yukawa coupling is sizeable.


\section{Results and collider bounds}\label{sec:coll}
In what follows, numerical results for the allowed parameters for each of the Higgs portal models introduced in Sec.~\ref{sec:models} will be shown. Besides constraints from the relic density and direct DM detection, we also evaluate bounds from current LHC data at $\sqrt{s}=8\tev$ (LHC8), as well as projections for upcoming LHC runs at $\sqrt{s}=14\tev$ (LHC14) and future planned colliders. In particular we consider the reach of a proton-proton collider with $\sqrt{s}=100\tev$ (called FCC-hh in the following) and of a high-energy $e^+e^-$ collider with $\sqrt{s}=1\tev$, such as the planned International Linear Collider (ILC).

In the singlet-singlet model, the experimental sensitivity to the new states is due to the Higgs-singlet mixing. This modifies the couplings of the SM-like Higgs, $h_1$, to the SM fermions and gauge bosons, which can be determined by measurements of the Higgs production rate in different decay channels. One can also search directly for the second scalar state, $h_2$, in the same channels as for the SM Higgs, but with a different mass and different production cross-section. The most sensitive channels are $h_2\to WW$ and $h_2\to ZZ$.

In the singlet-doublet and doublet-triplet models, the particle spectrum involving several neutral and charged fermions can lead to LHC signatures with leptons and missing energy. The most important channel is given by
\beq
\begin{aligned}
q\bar{q}' &\to W^{*-} \to \chi^-\chi_{m,h}^0, \\
\chi^- &\to \chi_l^0 W^{*-} \to \chi_l^0 \ell^- \bar{\nu}_\ell, 
\quad
\chi^0_{m,h} \to \chi_l^0 Z^{*} \to \chi_l^0 \ell^- \ell^+, \qquad 
(\ell = e,\mu)
\end{aligned}
\label{eq:lhcsign}
\eeq
and its charge-conjugated versions. The production process, mediated by an off-shell $W$ boson, is present both in the Majorana and Dirac DM models, and its rate is governed by the SU(2) gauge quantum numbers of the fermion multiplets. The only decay channel for the lightest charged fermion is also through an off-shell $W$ boson. For the next-to-lightest neutral fermion, tree-level decays can be mediated by an off-shell $Z$ or Higgs boson, but the latter is suppressed to a negligible level by the small Higgs-boson width.

The process \eqref{eq:lhcsign} leads to a final state of three leptons ($3\ell$) and missing energy, with a phenomenology similar to the neutralino--chargino
sector of the MSSM with heavy scalars. This signature has been searched for by the ATLAS and CMS experiments~\cite{Aad:2014nua,Khachatryan:2014qwa}. If the mass difference between the DM fermion and the heavier fermions is small, which is the case in the regions of parameter space where co-annihilation is important, the leptons can become too soft to pass the usual multi-lepton selection cuts. In this case, mono-jet searches provide the strongest limits from the published 8-TeV LHC data~\cite{Khachatryan:2014rra,Aad:2015zva}.

Notice that the heavier neutral fermions $\chi_{m,h}^0$ can in general also decay through a loop-induced radiative decay, $\chi_{m,h}^0 \to \chi^0_l + \gamma$. However, the branching fraction is suppressed to the percent level if either the heavy state, $\chi_{m,h}^0$, or the light state, $\chi^0_l$, has a dominant doublet component~\cite{Bramante:2014tba}, as is the case for the models considered here.

\subsection{Singlet-Singlet Model}\label{sec:coll-ss}
At colliders, this model can be most effectively tested by searching for the effects of the singlet scalar mediator, $S$. It mixes with the Higgs, $h$, so that the mass eigenstates $h_{1}$ and $h_2$ are admixtures of $S$ and $h$,
see Eq.~\eqref{eq:mixingSh}. As a result, the production and decays of the SM-like scalar $h_1$ to SM fermions and gauge bosons are suppressed by a common factor $\cos^2\alpha$, as discussed in Sec.~\ref{sec:gamgam}.

When interpreting the CMS and ATLAS Higgs measurements, we need to distinguish two cases. If $m_\chi<m_{h_1}/2$, then $h_1\to \chi\bar \chi$ is allowed, and thus the invisible branching ratio of the Higgs, $\text{Br}(h\to {\rm inv})$, could be sizeable. Allowing for an invisible decay of the Higgs and demanding $\kappa_V\leq1$, ATLAS obtains in a global analysis $\kappa_V>0.93$ at 95\% C.L., $\kappa_{f}=1.05\pm 0.16$ and Br$(h_1\to {\rm inv})<0.13$~\cite{ATLAS-CONF-2015-007}. (CMS obtains a much looser bound Br$(h_1\to {\rm inv})<0.49$ for $\kappa_V\leq1$~\cite{Khachatryan:2014jba}.) The invisible decay width is given by
\beq
\Gamma(h_1\to \chi \bar \chi)=\frac{m_{h_1}}{16 \pi}s_\alpha^2\bigg[y^2\Big(1-\frac{4m_\chi^2}{m_{h_1}^2}\Big)+y_5^2\bigg]\Big(1-\frac{4m_\chi^2}{m_{h_1}^2}\Big)^{1/2},
\eeq
and thus depends on $s_\alpha$, as well as on $y$ and $y_5$. The correct relic abundance can be obtained for $y$ as small as ${\mathcal O}(10^{-2})$~\cite{Esch:2013rta}. This means that, allowing for all values of $y$, $y_5$ that satisfy the Planck measurement \eqref{eq:cdm}, the strongest constraint on $c_\alpha$ is not the bound on Br$(h_1\to {\rm inv})$,
 but rather the bound on $\kappa_V$. We therefore have at $95\%$ C.L.
\beq\label{eq:calpha:bound}
c_\alpha>0.93\quad \text{for}\ m_\chi < m_{h_1}/2,
\eeq
or $s_\alpha<0.37$. 

If $m_\chi>m_{h_1}/2$, the invisible decay width of the Higgs is zero. We can therefore use the results of global fits of the Higgs data, where only $\kappa_{f}=\kappa_b=\kappa_\tau=\kappa_t$ and $\kappa_V=\kappa_W=\kappa_Z$ are varied. Combining the CMS results, $\kappa_V=1.01\pm0.07$ and $\kappa_{f}=0.87^{+0.14}_{-0.13}$~\cite{Khachatryan:2014jba}, and ATLAS results, $\kappa_V=1.09\pm0.07$, $\kappa_{f}=1.11^{+0.17}_{-0.15}$~\cite{ATLAS-CONF-2015-007}, gives $c_\alpha=1.035\pm0.045$ or a $95\%$ C.L. lower bound
\beq
c_\alpha>0.948\quad \text{for}\ m_\chi>m_{h_1}/2, \label{eq:bound:calpha}
\eeq
corresponding to $s_\alpha<0.318$. The bound \eqref{eq:bound:calpha} also applies for $m_\chi $ just slightly below $m_{h_1}/2$, where the invisible decay width of the Higgs is negligible because of phase space suppression.
The above results assume that $h_2$ is not (approximately) mass degenerate with $h_1$. In the remainder of this subsection we will assume that this is true and that, furthermore, $h_2$ is appreciably heavier than $h_1$. For bounds away from this limit in a subset of our parameter space, see for instance Ref.~\cite{Falkowski:2015iwa}.

The bounds on $\alpha$ can be improved by Higgs coupling measurements at future colliders~\cite{Englert:2014uua}. The LHC with $\sqrt{s}=14\tev$ and 3000~fb$^{-1}$ luminosity (HL-LHC) can establish a limit of $c^2_\alpha \gtrsim 0.9$ at $95\%$ C.L., assuming that the central values of all observables agree with the SM prediction. The projected reach of ILC with $\sqrt{s} \leq 500\gev$ and 500~fb$^{-1}$ is $c^2_\alpha \gtrsim 0.96$.

\begin{figure}
\makebox[.495\textwidth][l]{\hspace{1.3cm}\scriptsize $m_\chi < m_{h_2}/2$}\hfill%
\makebox[.495\textwidth][l]{\hspace{1.3cm}\scriptsize $m_\chi > m_{h_2}/2$}\\
\includegraphics[width=.495\textwidth]{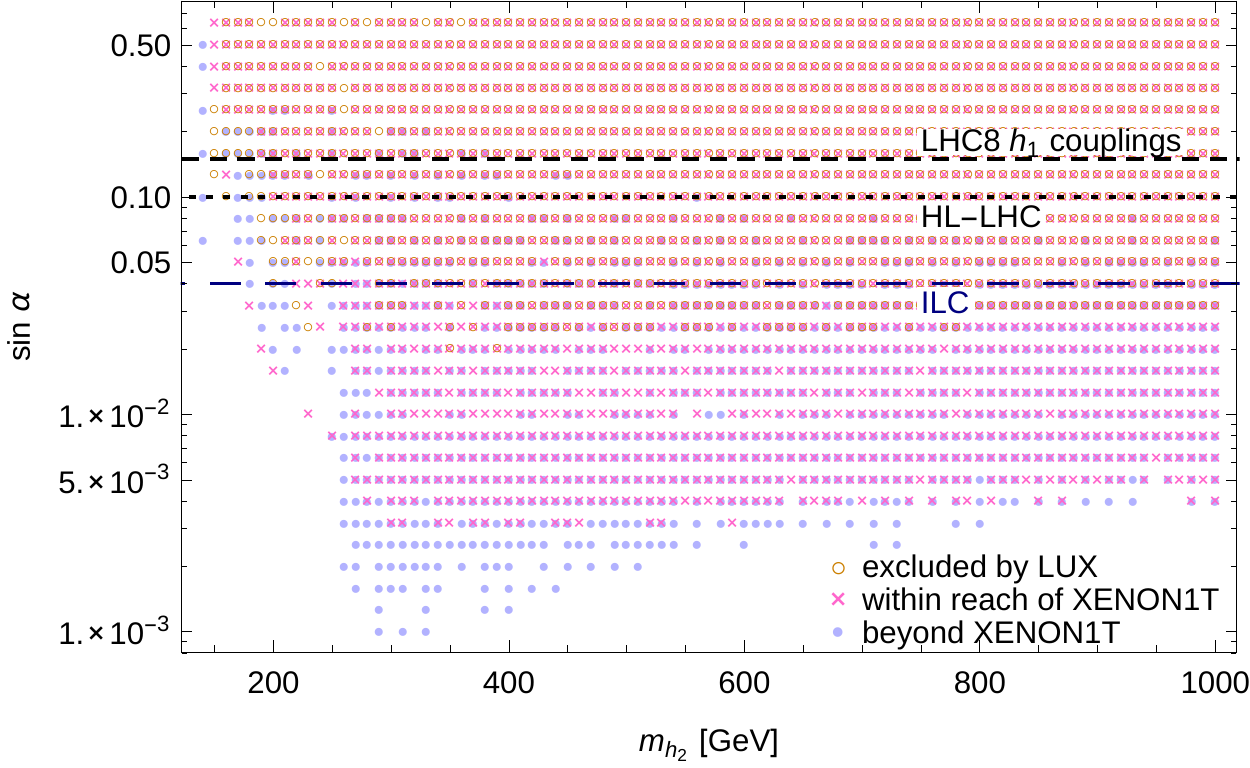}\hfill%
\includegraphics[width=.495\textwidth]{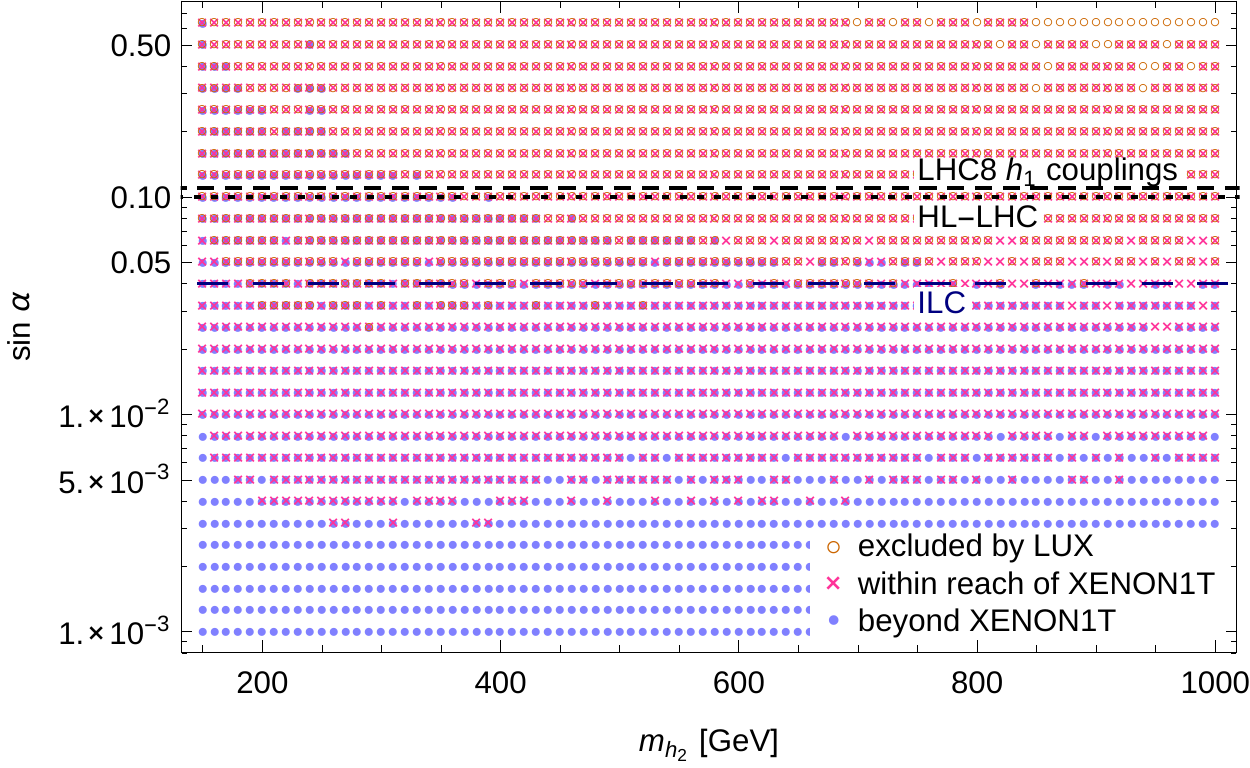}%
\vspace{-2ex}
\caption{Allowed parameter space for the singlet-singlet model consistent with the requirement that the thermal relic density of $\chi$ accounts for all dark matter in the universe, $\Omega_\chi = \Omega_{\rm DM}$. The different colors (shapes) of the points indicate current and future 90\% C.L.\ exclusions from direct detection experiments. Also shown are current 95\% C.L.\ limits from Higgs coupling measurements at the LHC (dashed line), and future projections for LHC14 with 3000~fb$^{-1}$ (dotted) and ILC with $\sqrt{s} \leq 500\gev$ (long dashed). The left panel corresponds to $m_\chi < m_{h_2}/2$, which forbids the annihilation channels $\chi\chi \to h_2h_{1,2}$, while in the right panel $m_\chi > m_{h_2}/2$.}
\label{fig:ssh}
\end{figure}
In Fig.~\ref{fig:ssh}, these bounds are compared to the region of parameter space of the singlet-singlet model that is consistent with the relic density constraint in Eq.~\eqref{eq:cdm}. The points in the figure are obtained through a parameter scan, where the model parameters are varied randomly within the following ranges:
\begin{align}
&100\gev < m_\chi < 1000\gev, &&-10<\lambda'<10, &&-1000\gev< \mu_S < 1000\gev.
\end{align}
The values of $\mu'$ and $m_S$ are determined from the mixing angle $s_\alpha$ and the masses $m_{h_{1,2}}$, see Eq.~\eqref{eq:mixingSh2}, while $\lambda_S$ is irrelevant for the relic density calculation. For simplicity and to avoid constraints from CP violation, we take $y_5=0$. For each random point, the value of $y$ is fixed by the relic density constraint, Eq.~\eqref{eq:cdm}, and the condition $y < 3$ is imposed to ensure perturbativity. Points shown as yellow circles are excluded by limits on the spin-independent direct detection cross-section from LUX~\cite{Akerib:2013tjd}, while the red crosses and blue dots are within and beyond the projected reach of XENON1T~\cite{Aprile:2012zx}, respectively.

One can also search directly for the heavy Higgs, $h_2$. The decay widths to SM fermions and gauge bosons are given by
\beq
\Gamma(h_2\to f\bar f)=s_\alpha^2 \Gamma(H\to f\bar f)_{\rm SM}, \qquad \Gamma(h_2\to VV)=s_\alpha^2 \Gamma(H\to VV)_{\rm SM},
\eeq
where $V=W,Z,\gamma, g$, and $\Gamma(H\to XX)_{\rm SM}$ is the partial width of the would-be SM Higgs if it had a mass $m_{h_2}$. If only the SM decay channels are open, the branching ratios of $h_2$ are thus given by the branching ratios of the SM Higgs with mass $m_{h_2}$. Similarly, the production cross-section is $\sigma(h_2)=s_\alpha^2 \sigma(H)_{SM}$, and is given by the would-be SM Higgs production with mass $m_{h_2}$.

For $m_{h_2}>2 m_{h_1}$, the decay $h_2\to h_1 h_1$ is kinematically allowed. It proceeds through the interactions
\beq
\begin{split}\label{eq:h2h1h1}
{\cal L}\supset\; & \mu^2\frac{h^3}{v}-\frac{\mu'}{2}S h^2-\lambda' v S^2 h - \mu_S S^3\\
=\;& h_2 h_1^2\Big[3 \frac{\mu^2}{v}s_\alpha c_\alpha^2-\frac{\mu'}{2}\big(c_\alpha^2- 2 s_\alpha^2\big) c_\alpha -\lambda' v\big(s_\alpha^2-2 c_\alpha^2\big) s_\alpha - 3 \mu_S c_\alpha s_\alpha^2\Big]+\cdots.
\end{split}
\eeq
If the mixing angle is small, $s_\alpha\ll 1$, then the $h_2h_1h_1$ coupling is equal to $\mu'/2$. This means that the $h_2\to h_1 h_1$ branching ratio can be large and will dominate over $h_2\to WW, 
ZZ, t\bar t$ for $\mu'\gg v$. On the other hand, if $\mu'\ll v$, the
decay $h_2\to h_1 h_1$ will be subleading. It is also possible to make the $h_2\to h_1 h_1$ branching ratio small while keeping $s_\alpha$ large, by canceling different 
contributions in Eq.~\eqref{eq:h2h1h1}. For $m_{h_2}>2 m_\chi$,
 the decay $h_2\to {\rm inv.}$ is open and can become dominant for sizeable values of $y$ or $y_5$. The actual constraint from $h_2$ decays thus strongly depends on the specific realization of the model.

\begin{figure}
\includegraphics[width=.495\textwidth]{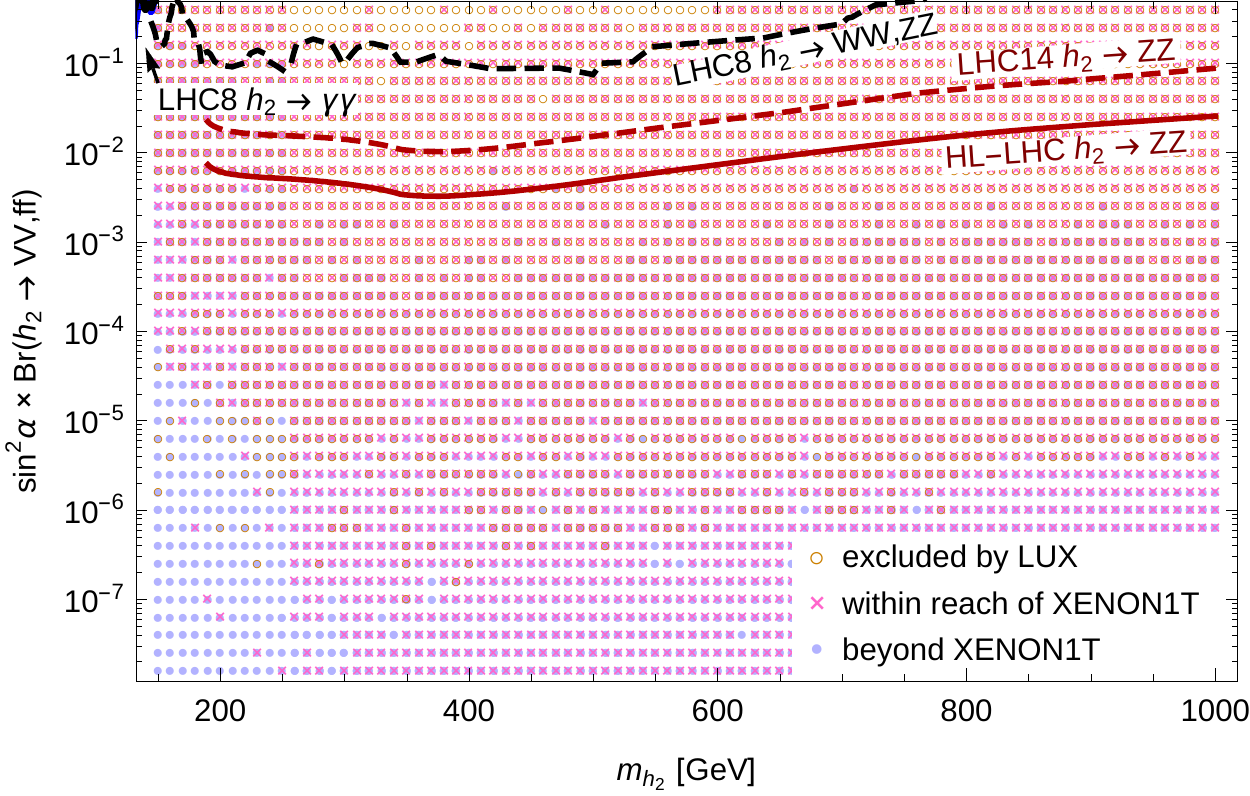}\hfill%
\includegraphics[width=.495\textwidth]{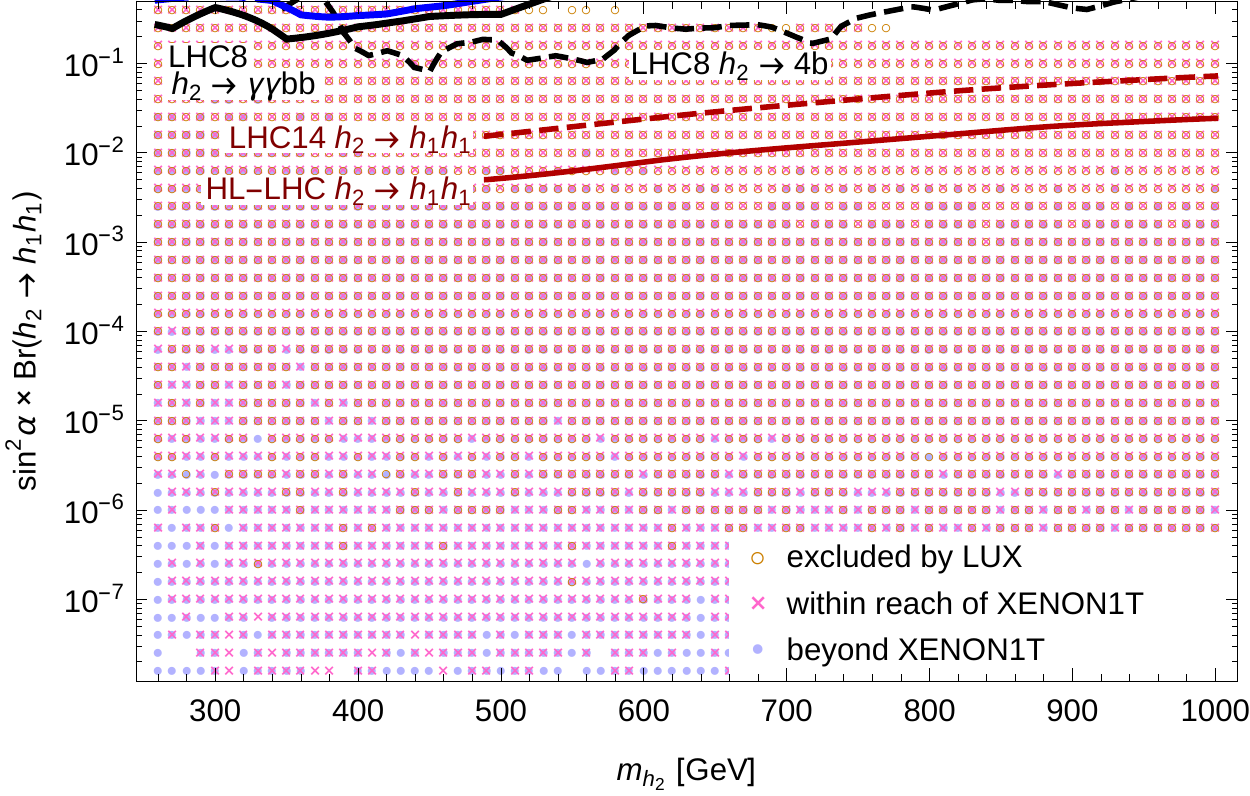}%
\vspace{-2ex}
\caption{Left: Existing bounds on $s_\alpha^2$ in the singlet-singlet model from $h_2\to \gamma\gamma$ (solid lines from ATLAS~\cite{Aad:2014ioa} and CMS~\cite{Khachatryan:2014ira}) and from $h_2\to WW, ZZ$ (dashed black line, CMS \cite{Khachatryan:2015cwa}) as a function of the heavy scalar mass, $m_{h_2}$. 
Right: Bounds on  $s_\alpha^2\times\text{Br}(h_2\to h_1 h_1)$ from $h_2\to h_1 h_1\to 2\gamma 2b$ (solid black from CMS \cite{CMS-PAS-HIG-13-032}, solid blue from ATLAS \cite{Aad:2014yja}) and from $h_2\to h_1 h_1\to 4b$ (dashed black from CMS \cite{CMS-PAS-HIG-14-013}). The projected exclusions at the 14-TeV LHC with 300~$\rm fb^{-1}$ (3000~$\rm fb^{-1}$) \cite{Buttazzo:2015bka} are shown as dashed (solid) red lines. The colored points indicate the parameter region consistent with the relic density constraint, $\Omega_\chi = \Omega_{\rm DM}$, with the different colors (shapes) denoting current and future 90\% C.L.\ exclusions from direct detection experiments.}
\label{fig:singlet-singlet}
\end{figure}

In Fig.~\ref{fig:singlet-singlet}~(left), we 
show the bounds from searches for direct decays of $h_2$ into SM final states. They can be expressed as a constraint
on $s_\alpha^2 \times \text{Br}(h_2\to VV,f\bar{f})$ as a function of $m_{h_2}$, where $V=W,Z,\gamma,g$. 
The most constraining channels are $h_2\to WW, ZZ$, with the resulting bound from CMS shown as a dashed black curve \cite{Khachatryan:2015cwa}. The searches in the di-photon channel, $h_2\to \gamma\gamma$, are effective at lower $h_2$ masses. The resulting bounds from ATLAS \cite{Aad:2014ioa} and CMS \cite{Khachatryan:2014ira} are shown as solid curves in the upper left corner.
 In Fig.~\ref{fig:singlet-singlet}~(right), we also show bounds on  $s_\alpha^2\times \text{Br}(h_2\to h_1 h_1)$  from di-Higgs searches. The CMS \cite{CMS-PAS-HIG-13-032} and ATLAS \cite{Aad:2014yja} bounds from $h_2\to h_1 h_1\to 2\gamma 2b$ are shown as solid black and blue lines, respectively.  
The CMS bound from $h_2\to h_1 h_1\to 4b$~\cite{CMS-PAS-HIG-14-013} is shown as a dashed black line. There are also CMS di-Higgs searches with di-photon and leptonic final states~\cite{Khachatryan:2014jya} and $h_2\to \tau\tau$~\cite{Khachatryan:2014wca}. However, with their current precision these results do not constrain $s_\alpha$. We see that relatively large mixing angles $s_\alpha\sim 0.3$ are allowed for all $h_2$ masses, comparable to the constraints from global Higgs coupling measurements in Eq.~\eqref{eq:calpha:bound}. In Fig.~\ref{fig:singlet-singlet} we also show as dashed (solid) red lines the projected exclusions at 14~TeV LHC with 300~$\rm fb^{-1}$ (3000~$\rm fb^{-1}$) obtained in Ref.~\cite{Buttazzo:2015bka}. At the end of the high luminosity LHC run, mixing angles as small as $s_\alpha\sim0.05$ can be probed for $m_{h_2}\sim 400$~GeV. Notice that for large $h_2$ masses, large mixings are additionally constrained by electroweak precision tests~\cite{Martin-Lozano:2015dja,Falkowski:2015iwa}.

At a 100 TeV collider, one can also search directly for DM production through an off-shell singlet mediator in the monojet signal, $pp\to h_1^*(\to \chi \bar \chi)j$, even if $h_1$ does not decay to DM \cite{Craig:2014lda}. This signature is quite challenging due to its very small cross-section, and can thus be observed only in a small parameter region with $m_\chi$ just above $m_{h_1}/2$. More promising is the signal with two jets and missing energy, where the presence of an additional jet allows for easier discrimination from the background~\cite{Khoze:2015sra}.

As can be seen from the figures, current and future direct detection experiments can cover a significantly larger portion of the parameter space of the singlet-singlet model than collider experiments. However, there are also many parameter points that are not constrained by available data from LUX, but that are excluded by LHC results for Higgs couplings and heavy Higgs searches. Therefore, collider and direct detection experiments provide complementary information for probing this model. On the other hand, the correct relic density can be realized for very small values of the mixing angle, $\sin\alpha \lesssim 0.01$, which cannot be tested conclusively either at colliders or through direct detection.

\subsection{Majorana Singlet-Doublet Model}
For $m_l^0 \gtrsim 100\gev$, co-annihilation is required in the Majorana singlet-doublet model to obtain the correct thermal relic density for weak Yukawa couplings, $y < 1$. This implies that the mass difference $m_m^0 - m_l^0$ is preferred to be a few tens of GeV. For DM masses in the TeV range the model thus becomes significantly fine-tuned. We therefore restrict ourselves to the range $m_l^0 < 1\tev$.

\begin{figure}
\includegraphics[width=.7\textwidth]{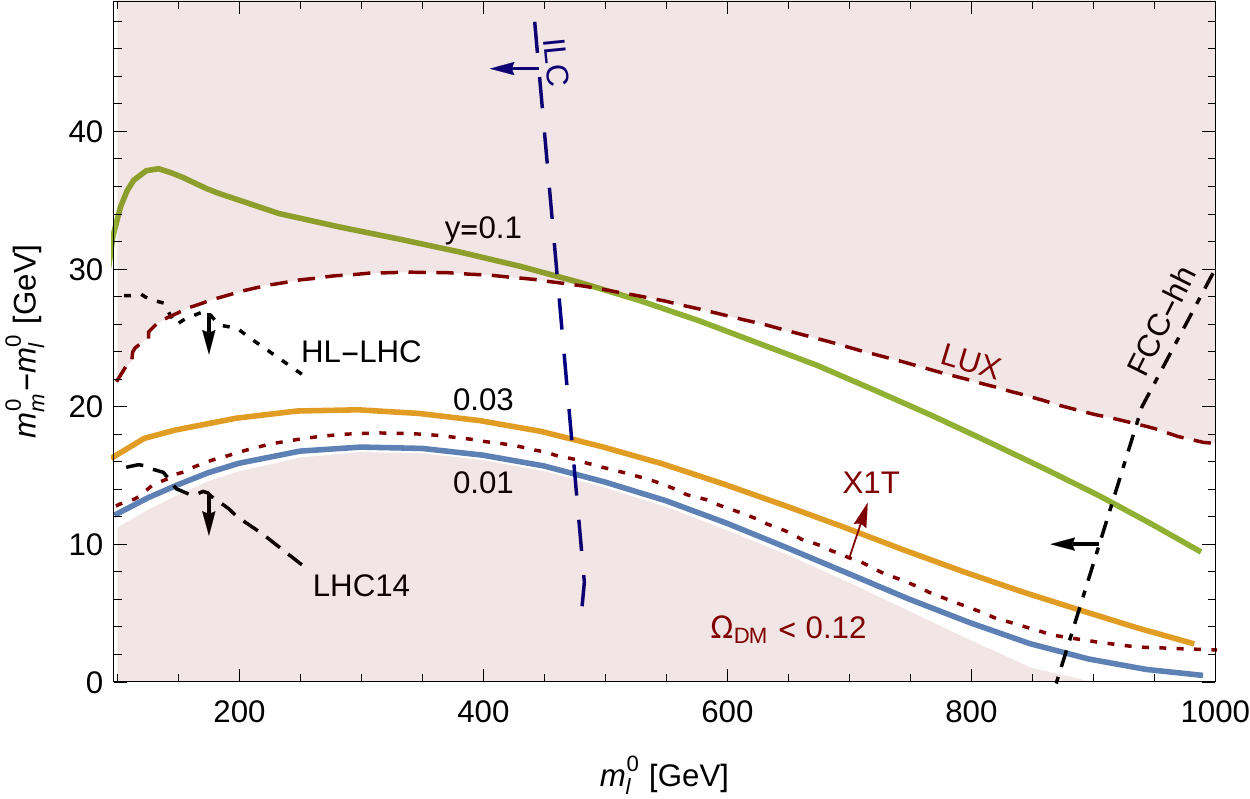}
\caption{Allowed parameter space for the Majorana singlet-doublet model, under the assumption that 
$\Omega_{\chi^0_l} = \Omega_{\rm DM}$. The solid lines indicate the correct DM density for different values of the Yukawa coupling $y$. The shaded regions are excluded by yielding too small of a relic density (bottom) and by direct detection limits from LUX (top, red dashed). Also shown are projected 95\% C.L.\ limits from jet plus soft leptons at LHC14 with 300~fb$^{-1}$ (short dashed) and 3000~fb$^{-1}$ (dotted) and FCC-hh with $\sqrt{s}=100\tev$ and 3000~fb$^{-1}$ (dot-dashed), as well as from ILC with $\sqrt{s}=1\tev$ (long dashed). The red dotted line depicts the expected 90\% C.L.\ direct detection limit from XENON1T. See text for details.}
\label{fig:majsd}
\end{figure}

If the mass difference becomes too small, the relic density constraint cannot be satisfied even for very small values of $y$. This is depicted by the shaded region at the bottom of Fig.~\ref{fig:majsd}. On the other hand, large values of $y$ are constrained by limits on the spin-independent direct detection cross-section from LUX \cite{Akerib:2013tjd} (see shaded region at the top of the plot). 

In the remaining allowed part of parameter space, the mass difference $m_m^0 - m_l^0$ is relatively small, so that the leptons from the process \eqref{eq:lhcsign} are soft and fail the selection cuts for the $3\ell$ signature~\cite{Aad:2014nua,Khachatryan:2014qwa}. On the other hand, the production cross-section for \eqref{eq:lhcsign} is too small to be constrained by the available mono-jet data \cite{Khachatryan:2014rra,Aad:2015zva}. As a consequence, no bound on the cosmologically preferred parameter space of the singlet-doublet model is obtained from LHC8 data (see also Refs.~\cite{Schwaller:2013baa,Martin:2014qra,Han:2014sya}).
Similarly, the projected reach for the $3\ell$ signal of the LHC14 \cite{CMS:2013xfa,ATLAS:2013hta} and FCC-hh \cite{Gori:2014oua,Acharya:2014pua} does not extend into the white region in Fig.~\ref{fig:majsd}. 

Instead, requiring a hard initial-state jet can help to trigger on events with soft leptons and improve the signal-to-background ratio for this case. Several authors have analyzed this signature, consisting of at least one hard jet, large missing energy, and at least two soft leptons, and found it to be promising for the parameter region preferred by co-annihilation~\cite{Gori:2013ala,Schwaller:2013baa,Low:2014cba,Han:2014kaa,Bramante:2014tba}. We have obtained the estimated 95\% C.L.\ reach of LHC14 by recasting the analysis of Ref.~\cite{Schwaller:2013baa}. Concretely, Ref.~\cite{Schwaller:2013baa} contains results for a Majorana singlet-triplet scenario, with the dominant signal contribution stemming from the process \eqref{eq:lhcsign}. In our case, since we have heavy doublet fermions instead of a triplet, the production cross-section is reduced by a factor of two. After accounting for this change in the production rate, the projected limits shown by the short-dashed (for 300 fb$^{-1}$) and dotted (for 3000 fb$^{-1}$) lines in Fig.~\ref{fig:majsd} are obtained. As can be seen from the figure, such an analysis will only be able to test this model for relatively small DM masses, $m_l^0 \lesssim 250\gev$, and only in the high-luminosity run of the LHC.

However, a much larger part of the parameter space can be probed
with FCC-hh with $\sqrt{s}=100\tev$. We use the results for higgsino and higgsino-bino scenarios in Ref.~\cite{Low:2014cba}, which directly correspond to our Majorana singlet-doublet model. The authors of Ref.~\cite{Low:2014cba} have simulated the signals and backgrounds for the monojet signature and the jet plus soft leptons signature. The projected 95\% C.L.\ limit is shown by the dash-dotted line in Fig.~\ref{fig:majsd}. We find that the astrophysically favored parameter space for $m_l^0 \lesssim 900\gev$ can be covered.

Also shown in the figure is the projected reach of the ILC with $\sqrt{s}=1\tev$, which has been evaluated by extrapolating the results of Ref.~\cite{Berggren:2013bua}. Due to its clean environment and low backgrounds, the ILC will be able to probe this model for masses of the doublet fermions, $m_m^0 = m^+ = m_D$, up to almost half of the center-of-mass energy (see long-dashed line in Fig.~\ref{fig:majsd}).

In addition to constraints from future collider experiments, the allowed parameter space of the Majorana singlet-doublet model will also be probed by upcoming direct detection experiments. The dotted red line in Fig.~\ref{fig:majsd} indicates the projected reach of the XENON1T experiment \cite{Aprile:2012zx}. As evident from the figure, it will be able to cover large parts of the parameter space, except for very small Yukawa couplings, $y \lesssim 0.02$.

\begin{figure}
\includegraphics[width=.7\textwidth]{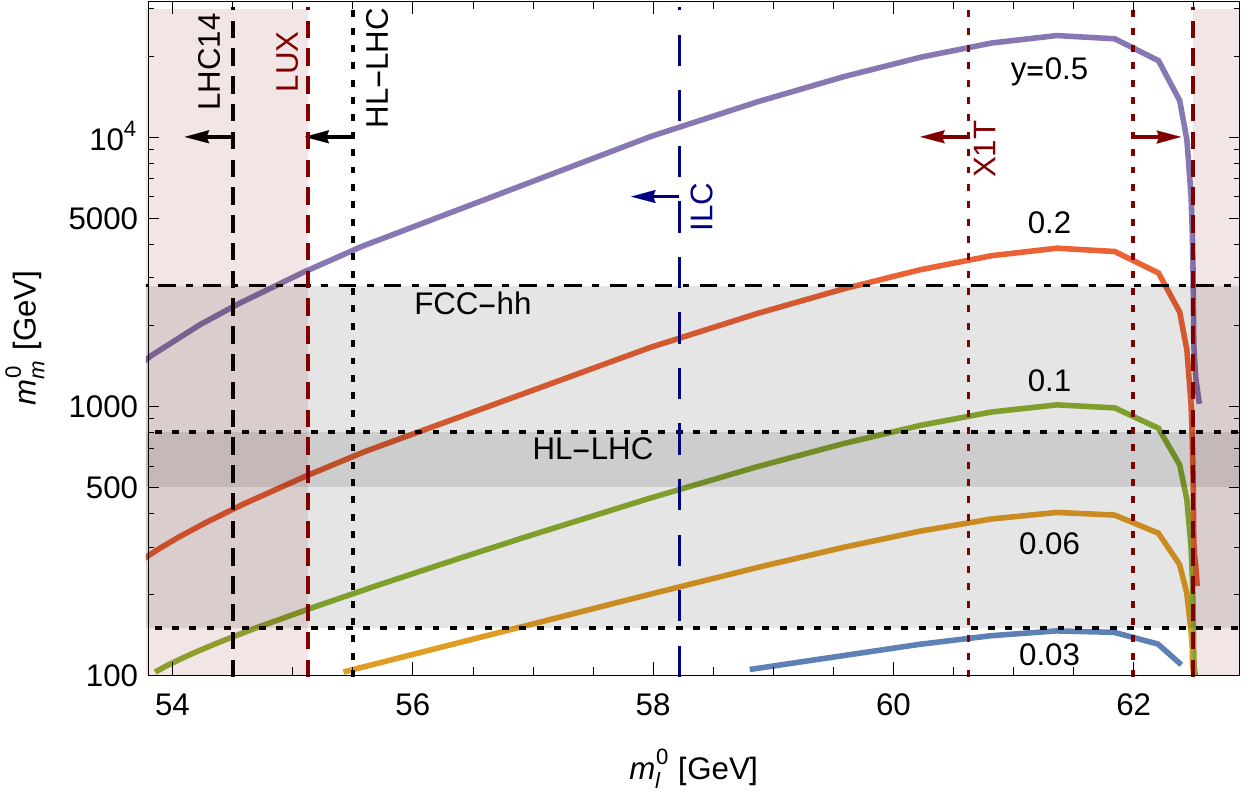}
\caption{Allowed parameter space for the Majorana singlet-doublet model in the Higgs resonance region, with the constraint $\Omega_{\chi^0_l} = \Omega_{\rm DM}$. The solid lines indicate the correct DM density for different values of the Yukawa coupling $y$. The red-shaded region is excluded by direct detection limits from LUX. Also shown are projected 95\% C.L.\ limits from 3$\ell$ searches at LHC14 with 3000~fb$^{-1}$ (dotted horizontal lines and shading) and FCC-hh with 3000~fb$^{-1}$ (dot-dashed horizontal line and shading). The vertical lines depict the projected limits from $h \to \text{invisible}$ measurements at LHC14 with 300~fb$^{-1}$ (short dashed) and 3000~fb$^{-1}$ (dotted), and at ILC with $\sqrt{s}=250\gev$ (long dashed), as well as from direct detection searches at XENON1T (red dotted). See text for details.}
\label{fig:majsdh}
\end{figure}

\vspace{\bigskipamount}
The correct relic density in this model can also be achieved by annihilation of $\chi^0_l$ pairs through the Higgs resonance. In this case, co-annihilation does not play any role, and the other fermion states can be much heavier than the DM state, as shown in Fig.~\ref{fig:majsdh}. 
In this scenario, both the annihilation and the direct detection cross-sections are mediated by Higgs exchange, so that the direct detection bound from LUX \cite{Akerib:2013tjd} becomes independent of the heavy fermion mass in this plot. It excludes part of the parameter space that produces the correct relic density. 

The remaining parameter space can be probed in three ways. The $3\ell$ signature from the process \eqref{eq:lhcsign} can be observed if the mass difference $m_m^0 - m_l^0$ is sufficiently large, but the heavy states $\chi_m^0$ and $\chi^\pm$ are not too heavy. The first condition ensures that the leptons can pass the trigger and selection requirements, while the second is related to the need for a large enough signal production cross-section. For LHC14 with 3000~fb$^{-1}$, we estimate its 95\% C.L.\ exclusion limit for this signature based on simulation results from the ATLAS collaboration \cite{ATLAS:2013hta}. To account for the smaller cross-section in the Majorana singlet-doublet model compared to the supersymmetric singlet-triplet scenario studied by ATLAS, we conservatively take the $5\sigma$ rather than the 95\% C.L.\ contour from Fig.~10 in Ref.~\cite{ATLAS:2013hta}.\footnote{While this choice will underestimate the reach of LHC14, we note that the covered parameter space does not change much between the 95\% C.L.\ and $5\sigma$ contours from Fig.~10 in Ref.~\cite{ATLAS:2013hta}.} For FCC-hh with $\sqrt{s}=100\tev$ and 3000~fb$^{-1}$, we adopt the results from Sec. III.C in Ref.~\cite{Gori:2014oua}. As evident from the horizontal bands in Fig.~\ref{fig:majsdh}, the combination of LHC14 and FCC-hh results can cover most of the astrophysically allowed parameter space, except for relatively large values of the Yukawa coupling $y$.

Alternatively, this scenario can be tested by measurements of the invisible Higgs width. For $\chi^0_l < m_h/2$, the Higgs boson can decay into pairs of DM particles, $h \to \chi^0_l\chi^0_l$, which escape undetected. Future LHC data will be able to put strong limits on the invisible branching fraction of 17\% with 300~fb$^{-1}$ and 6\% with 3000~fb$^{-1}$, under somewhat favorable assumptions for the systematic errors~\cite{CMS:2013xfa}. A much more precise constraint of Br$(h\rightarrow \rm{inv})<0.3\%$ is expected from ILC data taken at $\sqrt{s}=250\gev$ (see Tab.~2.6 in Ref.~\cite{Baer:2013cma}).
The corresponding bounds on the parameter space of the model are depicted by the vertical lines in Fig.~\ref{fig:majsdh}. These bounds do not depend on the heavy fermion mass, $m_m^0$, since both the thermal relic density and the Higgs invisible width are governed by the same $h\chi^0_l\chi^0_l$ coupling.

Finally, future direct detection searches by XENON1T~\cite{Aprile:2012zx} will probe most of the allowed parameter space, as depicted by the red dotted lines in Fig.~\ref{fig:majsdh}. Combining the projected limits from collider and direct detection experiments, only the parameter region where the DM mass is very close to half the Higgs mass, $60.5\gev \lesssim m_l^0 \lesssim 62\gev$, and the mediator mass is very large, $m_m^0 \gtrsim 3\tev$, will remain inaccessible.

\subsection{Dirac Singlet-Doublet Model}
\begin{figure}
\includegraphics[width=.7\textwidth]{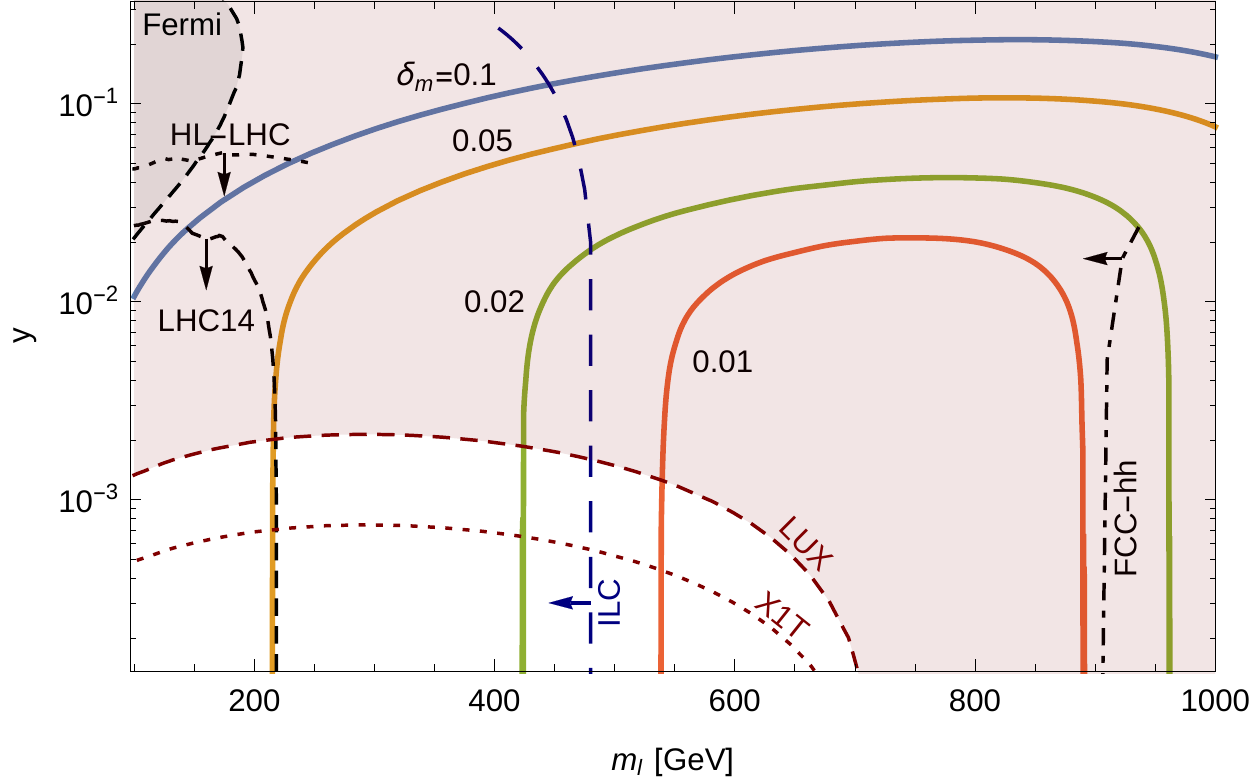}
\caption{Allowed parameter space for the Dirac singlet-doublet model, with the constraint $\Omega_{\chi^0_l} = \Omega_{\rm DM}$. The solid lines indicate the correct DM density for different values of the relative mass difference $\delta_m \equiv  (m_h^0-m_l^0)/m_l^0$. The red-shaded region is excluded by direct detection limits from LUX. Also shown are projected 95\% C.L.\ limits from jet plus soft leptons searches at LHC14 with 300~fb$^{-1}$ (short dashed) and 3000~fb$^{-1}$ (dotted), and at FCC-hh with 3000~fb$^{-1}$ (dot-dashed). The reach of ILC with $\sqrt{s}=1\tev$ is depicted by the long-dashed line, while the red dotted line indicates the expected 90\% C.L.\ limit from XENON1T. The gray region is excluded by the Fermi-LAT indirect DM searches.}
\label{fig:dirsd}
\end{figure}

As for the Majorana case, co-annihilation in the Dirac singlet-doublet model is active for most of the parameter points that satisfy the relic density constraint. In fact, the predicted value for $\Omega_{\chi_l^0}$ in this model is very sensitive to the mass difference $m_h^0-m_l^0$. Therefore, instead of displaying our results in the plane of $m_l^0$ and $m_h^0-m_l^0$, we chose $m_l^0$ and the Yukawa coupling $y$ as independent variables in Fig.~\ref{fig:dirsd}.

As expected, large regions of the parameter space are excluded by direct detection limits from LUX \cite{Akerib:2013tjd} (shaded region in plot), since the doublet component of $\chi_l^0$ couples to the $Z$ boson and thus leads to sizeable DM-nucleon interactions. In the remaining part of the parameter space, the mass difference $m_h^0-m_l^0$ is small, which is illustrated by the colored solid curves in Fig.~\ref{fig:dirsd}. This bound will be improved by future results from XENON1T \cite{Aprile:2012zx} (red dotted line in plot), but the region with small mass differences and small Yukawa couplings will remain difficult to probe by direct detection experiments. The upper left corner of the parameter space in Fig.~\ref{fig:dirsd} with large $y$ and small $m_l^0$ is also excluded by indirect searches. The 95\% C.L. bound from the latest analysis of dwarf spheroidal galaxies by the Fermi satellite~\cite{Ackermann:2015zua} is shown as a long-dashed line. The gray region to the left of the line is excluded.

As explained in the previous section, scenarios with such small mass differences can be best searched for by using the hard jet plus soft leptons signature at the LHC. The expected 95\% C.L.\ reach of LHC14 with 300~fb$^{-1}$ and 3000~fb$^{-1}$ from the analysis in Ref.~\cite{Schwaller:2013baa} is shown by the dashed and dotted lines in Fig.~\ref{fig:dirsd}, respectively. These bounds have been obtained by rescaling the results of Ref.~\cite{Schwaller:2013baa} to account for the production cross-section of $pp \to \chi^-\chi^0_h, \, \chi^+\bar{\chi}^0_h$ in the Dirac singlet-doublet model. Also shown are the projected reach of FCC-hh for the jet plus soft leptons signal from the analysis of Ref.~\cite{Low:2014cba} (dash-dotted line), as well as the reach of ILC with $\sqrt{s}=1\tev$ (long-dashed line).

As is evident from the figure, LHC14 will be able to probe part of the astrophysically viable parameter space, but only for light DM with $m_l^0 \lesssim 250\gev$. In contrast, ILC with $\sqrt{s}=1\tev$ can extend this reach to about $m_l^0 \lesssim 490\gev$, while FCC-hh is expected to cover the entire allowed parameter region.

\subsection{Majorana Doublet-Triplet Model}
In this subsection, we explore the parameter region of the Majorana doublet-triplet model for a pure doublet DM candidate, corresponding to the left panel of Fig.~\ref{fig:majdtx}. This scenario is characterized by sizeable mass splittings between different dark sector fermion states, so that co-annihilation is not important.

\begin{figure}
\includegraphics[width=.5\textwidth]{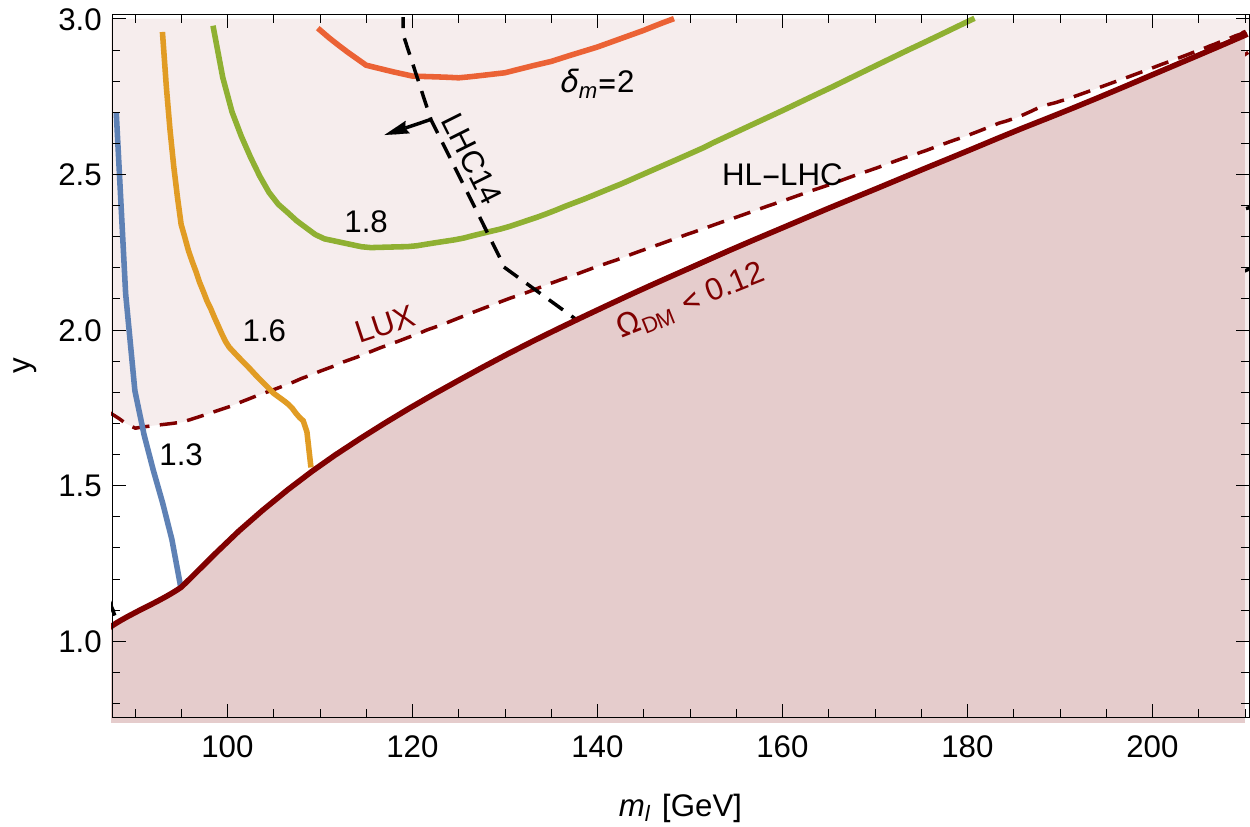}%
\includegraphics[width=.5\textwidth]{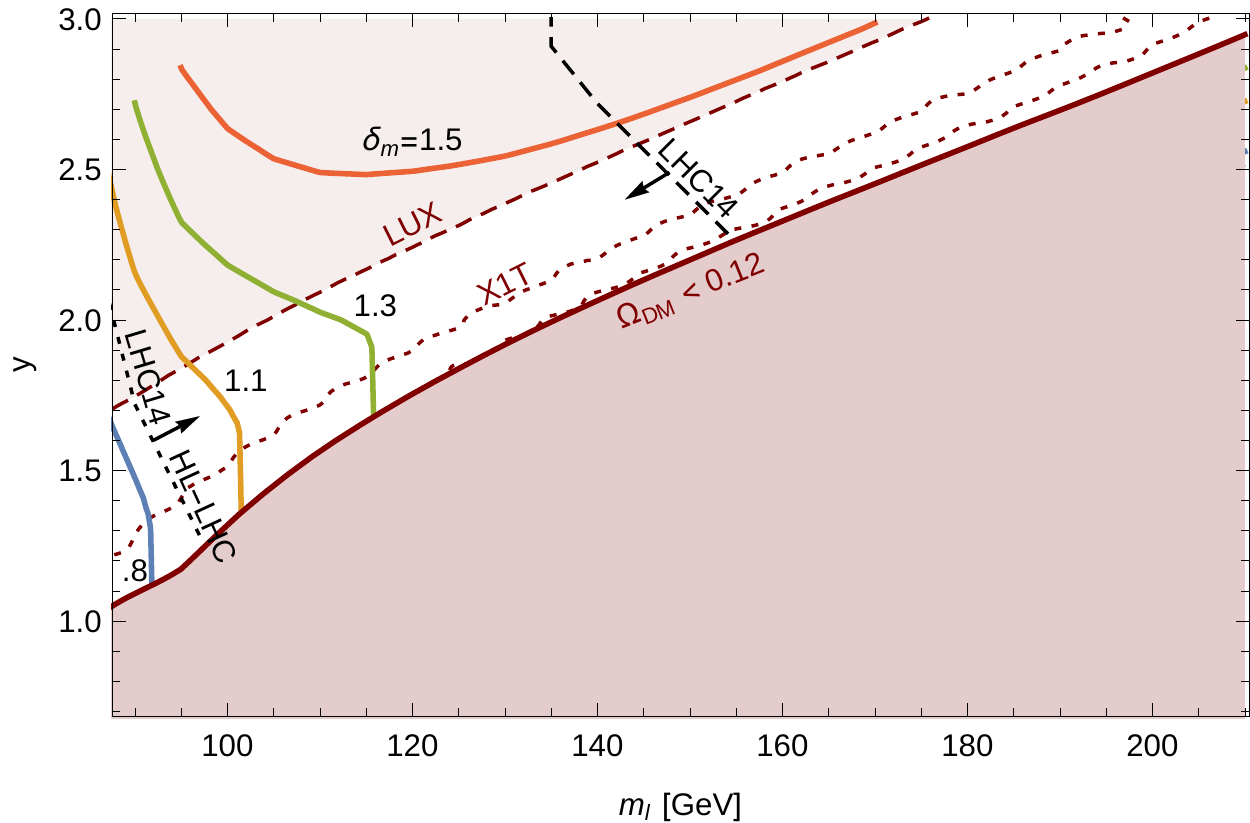}%
\caption{Allowed parameter space for the Majorana doublet-triplet model, with the constraint $\Omega_{\chi^0_l} = \Omega_{\rm DM}$. Each point in the $m_l^0$--$y$ parameter plane has two solutions for $\delta_m \equiv  (m_m^0-m_l^0)/m_l^0$, which are shown in the two panels. The red-shaded region at the bottom right is excluded by yielding too small of a relic density, while the light shaded regions at the top are constrained by data from LUX. Also shown are projected 95\% C.L.\ limits from $3\ell$ searches at LHC14 with 300~fb$^{-1}$ (short dashed) and 3000~fb$^{-1}$ (dotted), and the 90\% C.L.\ reach of XENON1T (red dotted). In the left panel, the entire allowed parameter region can be covered by LHC with 3000~fb$^{-1}$, as well as by XENON1T. Notice that the whole viable parameter space in both panels is already excluded by $h\to\gamma\gamma$ data from LHC8, assuming that there are no new fields coupling to the Higgs boson beyond the doublet and triplet fermions.}
\label{fig:majdt}
\end{figure}

For this model, it is not possible to display our results in the plane of $m_l^0$ and $m_m^0-m_l^0$, since for some values of $m_l^0$ and $m_m^0$ there are more than two possible values of $y$ that satisfy the relic density constraint. Instead, as in the previous subsection, $m_l^0$ and the Yukawa coupling $y$ are used as independent variables. Even in this case, there is a two-fold ambiguity for the value of $m_T$ for each point in the $m_l^0$--$y$ plane, as can be seen in Fig.~\ref{fig:majdtx}~(left). These two solutions are shown in the two panels in Fig.~\ref{fig:majdt}. The solid colored curves indicate the correct relic density for different values of the relative mass difference $\delta_m \equiv (m_m^0-m_l^0)/m_l^0$. Notice that large values of $y$ are constrained by vacuum stability~\cite{Dedes:2014hga}.

The region with small Yukawa couplings is excluded by the relic density constraint, since the mass difference between the two doublet-dominated neutral fermion states is relatively small in this case, leading to efficient $\chi^0_l$--$\chi^0_l$ annihilation. However, viable parameter points are obtained for larger Yukawa couplings, $1 \lesssim y < 3$, where the upper bound is imposed to ensure perturbativity, and for masses $m_l^0 < 220\gev$.

As already mentioned in Sec.~\ref{sec:direct-detection}, the DM-nucleon cross-section is loop-suppressed in this scenario. Nevertheless, due to the large values of the Yukawa coupling $y$ in the viable parameter space, there are still important constraints from direct detection experiments. The bounds from LUX~\cite{Akerib:2013tjd} are shown in Fig.~\ref{fig:majdt}. With the improved sensitivity of XENON1T~\cite{Aprile:2012zx}, almost the entire parameter space of the model can be covered.

At the LHC, the heavier neutral and charged fermions can be produced according to \eqref{eq:lhcsign}, leading to a $3\ell$ signal with missing energy. Since the mass differences are relatively large in the cosmologically viable parameter region, no hard initial-state jet is required. Both ATLAS and CMS have published limits on the production cross-section from $3\ell$ searches \cite{Aad:2014nua,Khachatryan:2014qwa} (see also additional material available in Ref.~\cite{Aad:2014nuaSupp}), which can be compared to the cross-section for the Majorana doublet-triplet model computed with {\tt CalcHEP}. It is found that the current data from ATLAS and CMS does not lead to any limit in the parameter region shown in Fig.~\ref{fig:majdt}.

Additionally, this model is constrained by LHC data on the branching fraction of the Higgs boson to two photons, as discussed in Sec.~\ref{sec:gamgam}. Since we require relatively light masses and large Yukawa couplings of the new fermions to obtain the correct relic density, the correction to $R_\gamma$ is sizeable. In fact, one finds $R_\gamma < 0.6$ for the whole viable parameter region shown in Fig.~\ref{fig:majdt} (see also Ref.~\cite{Dedes:2014hga}). This is in conflict with results from ATLAS and CMS yielding $R_\gamma = 1.17 \pm 0.27$~\cite{Aad:2014eha} and $R_\gamma = 1.14^{+0.26}_{-0.23}$~\cite{Khachatryan:2014ira} at the 68\% C.L., respectively. Assuming that the experimental uncertainties are Gaussian distributed, one thus finds that the Majorana doublet-triplet model with pure doublet dark matter is excluded at more than 95\% C.L..

However, it is worth pointing out that the $h\to\gamma\gamma$ branching fraction may be modified by other new physics unrelated to the DM sector, so that the Majorana doublet-triplet model may remain viable as part of a larger  theory. In this case, the model can be probed more robustly through direct searches for the dark sector fermions with future LHC data. For the projected reach of $3\ell$ searches, we use Fig.~10 in Ref.~\cite{ATLAS:2013hta}. To account for the smaller production cross-section due to the mixing between the dark fermions, we conservatively use the $5\sigma$ contour in that figure in lieu of the 95\% C.L.\ bound. 
Using these results, the estimated reach of LHC14 with 300~fb$^{-1}$ is illustrated by the short-dashed lines in Fig.~\ref{fig:majdt}.
For 3000~fb$^{-1}$, we find that the entire viable parameter region in the left panel of Fig.~\ref{fig:majdt} can be excluded with $3\ell$ data, while only a small corner in the lower left part of the right panel remains.


\section{Conclusions}
When probing the fermionic dark matter Higgs portal at colliders, the mediators cannot be integrated out, but rather must be considered as dynamical degrees of freedom. The inclusion of new signatures due to the on-shell production of the mediators can substantially extend the reach of collider searches for Higgs portal models. Using the example of three simple UV completions of the Higgs portal with scalar and fermion mediators, we have demonstrated the complementarity of direct detection experiments and high-energy colliders.

In the case of a scalar mediator, there is no direct connection between direct detection and collider observables, due to a large number of free parameters. For a singlet scalar, the observed thermal relic density can be accommodated with small Higgs-singlet mixing, especially if DM is heavier than the heavy scalar. Small mixing helps to evade direct detection bounds, as well as collider bounds from Higgs coupling measurements and direct searches for the heavier scalar. Future colliders, i.e. the HL-LHC, ILC or FCC-hh, and direct detection experiments can cover a large part of the model's parameter space. They will, however, not be able to test it conclusively.

In models with fermion mediators, obtaining the correct relic density requires either large Yukawa couplings or co-annihilation. The scenario with large Yukawa couplings is generally strongly constrained by direct detection, as are models with unsuppressed couplings to the $Z$ boson. The co-annihilation scenario is viable for small Yukawa couplings $y\lesssim 0.1$. Since it implies small mass splittings among the lightest states, dedicated search strategies are needed at colliders. These compressed spectra can be searched for in mono-jet signatures dressed with soft leptons, which are produced in the decay chain of the mediators.

For singlet DM and a doublet fermion mediator, several scenarios can satisfy all current bounds. If DM is a Majorana fermion, $Z$-mediated DM-nuclei interactions are absent and the co-annihilation scenario is viable for DM--mediator mass differences of about $10-30\gev$. While the LHC can access DM masses below $300\gev$, the scenario can be conclusively tested with XENON1T and future colliders up to TeV-scale DM masses. In the Higgs resonance region, $m_\chi \sim m_h/2$, the correct relic density can be obtained without co-annihilation for Yukawa couplings $y\sim 0.1$, if mediators are below the TeV scale. The LHC (and future hadron colliders) can probe mediators up to (beyond) the TeV scale with three-lepton signatures, leaving only a small corner of parameter space that neither colliders nor XENON1T can reach.

If, in turn, DM is a Dirac fermion, the singlet-doublet model is viable only for tiny Yukawa couplings $y\lesssim 10^{-3}$, where $Z$-mediated DM-nuclei interactions are sufficiently suppressed to evade LUX bounds. XENON1T can probe even smaller couplings. The LHC will test the region $m_\chi \lesssim 200\gev$ for arbitrarily small Yukawa couplings. The ILC and FCC-hh can cover higher masses and ultimately test the model conclusively.

In a model with one doublet and one triplet fermion field, which both carry weak charges, the only viable scenario is pure-doublet DM, where $Z$- and Higgs-mediated DM-nuclei interactions are absent at tree level. Despite the suppression, loop-induced scattering still leads to significant constraints from direct detection searches. In addition, the model in its minimal version is already excluded by the present Higgs property measurements at the LHC, due to the contributions of charged fermions to the Higgs decay to two photons. If these can be canceled by a new contribution unrelated to the dark sector, the extended model is viable for large Yukawa couplings $y > 1$ and DM masses below $200\gev$. XENON1T and direct searches for electroweak particles at LHC14 can cover almost the entire viable parameter space of this model.

Since the mediators in our models are generally not much heavier than the DM particles, an effective Higgs portal description is not useful. Exceptions are the Higgs resonance region in the Majorana singlet-doublet model and parts of the parameter space in the singlet-singlet model, where the DM--mediator mass splitting can be sizeable.

In summary, collider searches and direct detection experiments can pin down models with scalar mediators that have moderate mixing with the Higgs boson, and can fully explore models with fermion mediators, apart from small regions of parameter space. This is possible due to the combination of different experiments. The complementarity of probes will become even more important in the case of a discovery.

\section{Acknowledgments}
We thank Athanasios Dedes and Dimitrios Karamitros for useful discussions and comparison of results, Matthew McCollough for helpful comments, and Pedro Schwaller and Jose Zurita for sharing their numerical results of Ref.~\cite{Schwaller:2013baa} with us.
 A.F. and S.W. are supported in part by the U.S. National Science Foundation, grant PHY-1212635. J.Z. is supported in part by the U.S. National Science Foundation under CAREER Grant PHY-1151392.


\appendix

\section{Loop-induced Higgs boson interactions with Majorana DM}\label{app:one-loop-dd}
In this appendix, we give analytic formulae for the loop-induced Higgs coupling to DM in the Majorana doublet-triplet model with doublet-like DM, $\chi_{\ell}^0=\chi_b^0$ (see Sec.~\ref{sec:doublet:triplet}). The relevant term in the Lagrangian can be written as
\begin{equation}
\mathcal{\widehat{L}} = - \frac{\delta y}{2}\, \chi_b^0\chi_b^0\,h + {\rm h.c.}.
\end{equation}
We have calculated the Yukawa correction $\delta y$ at the one-loop level in unitarity gauge in the limit of zero momentum transfer to the Higgs boson. The result is given by
\begin{align}
\delta y & = \sum_{i,j=a,c}\,\Big[g_{ij,0}^h g_{ib}^Zg_{jb}^{Z\ast} F(Z,\chi_i^0,\chi_j^0) + 2\,g_{ij,+}^h g_{ib}^Wg_{jb}^{W\ast} F(W,\chi_i^+,\chi_j^+) \Big]\\\nonumber
 &\ \ \, + \sum_{i=a,c} \Big[ \frac{g}{c_W} |g_{ib}^Z|^2 F(Z,Z,\chi_i^0) + 2\,g |g_{ib}^W|^2 F(W,W,\chi_i^+) \Big],
\end{align}
where the couplings $g_{ij,0}^h$ etc. are listed in Tab.~\ref{tab:sm-dm-couplings}.
\begin{table}[tb]
\centering
\begin{tabular}{ccccc|cc}
\hline\hline
$j$ & $g_{jb}^Z$ & $g_{jb}^W$ & $g_{jj,0}^h$ & $g_{jj,+}^h$ & $k$ & $g_{ac,k}^h$ \\
\hline
$a$ & $+i\frac{g}{2c_W}\sin\theta_2$ & $+i\frac{g}{2}\sin\theta_2$ & $-\frac{y}{\sqrt{2}}\sin(2\theta_2)$ & $-\frac{y}{\sqrt{2}}\sin(2\theta_2)$ & $+$ & $-\frac{y}{\sqrt{2}}\cos(2\theta_2)$\\
$\ c\ $ & $-i\frac{g}{2c_W}\cos\theta_2$ & $+i\frac{g}{2}\cos\theta_2$ & $+\frac{y}{\sqrt{2}}\sin(2\theta_2)$ & $+\frac{y}{\sqrt{2}}\sin(2\theta_2)$ & $\ 0\ $ & $+\frac{y}{\sqrt{2}}\cos(2\theta_2)$\\
\hline\hline
\end{tabular}
\caption{Dark fermion couplings to SM bosons in the Majorana doublet-triplet model.}
\label{tab:sm-dm-couplings}
\end{table}
 The loop functions can be expressed as
\begin{align}
F(B,\chi_i,\chi_j) & = \frac{1}{16\pi^2}\frac{1}{m_B^2}\Bigg\{ \frac{m_i+m_j}{2m_D}\big(2m_B^2+(m_i-m_D)(m_j-m_D)\big)B_0(0,m_i^2,m_j^2) - A_0(m_i^2) - 2m_B^2\\\nonumber
& \hspace{2.2cm} + \frac{m_i+m_D}{2m_D}(m_i^2-m_B^2)B_0(0, m_i^2, m_B^2) + \frac{m_j-m_D}{2m_D}(m_j^2-m_B^2) B_0(0, m_j^2, m_B^2)\\\nonumber
& \hspace{2.2cm} + \Big[\frac{(m_i^2-m_D^2)^2+(m_i^2+m_D^2-2m_B^2-6m_im_D)m_B^2}{2m_D(m_j-m_i)}\,B_0(m_D^2, m_i^2, m_B^2) + (i\leftrightarrow j)\Big]\Bigg\},
\end{align}
\begin{align}
F(B,B,\chi_i) & = \frac{1}{16\pi^2}\frac{1}{m_B^2}\Bigg\{ \frac{m_B^3}{m_D}\frac{2m_D^2+m_im_D-m_i^2-2m_B^2}{(m_i+m_D)^2-m_B^2}\,B_0(0, m_B^2, m_B^2)\\\nonumber
&\hspace{1.3cm} + \frac{m_B}{m_D}\frac{(m_i+m_D)^2+2m_B^2}{(m_i+m_D)^2-m_B^2}\Big[m_i^2B_0(0,m_i^2,m_i^2)+(m_i^2+m_D^2-m_B^2)\Big]\\\nonumber
&\hspace{1.3cm} + \frac{m_i}{2m_Dm_B}(m_i-m_D)(m_i^2-m_B^2)B_0(0,m_i^2,m_B^2) + \frac{m_D - m_i}{2m_B}\big(A_0(m_i^2) + A_0(m_B^2)\big)\\\nonumber
&\hspace{1.3cm} - \frac{1}{2m_Dm_B}\Big[m_B^4\big((m_i+m_D)^2+2(m_i^2+m_D^2)-4m_B^2\big) + 2m_B^2m_im_D(m_i+m_D)^2\\\nonumber
&\hspace{1.3cm} + m_i^4(m_i^2+2m_im_D-m_D^2)+m_D^4(m_D^2+2m_im_D-m_i^2) - 4m_i^3m_D^3\Big]\frac{B_0(m_D^2, m_i^2, m_B^2)}{(m_i+m_D)^2-m_B^2}\Bigg\},
\end{align}
where $m_D=m_b^0$ is the doublet DM mass. The notation for the loop integrals $A_0$ and $B_0$ has been adopted from Ref.~\cite{Denner:1991kt}. Since $\chi_b^0$ self-energy corrections do not contribute to the $h\chi_b^0\chi_b^0$ vertex at one-loop, the result for $\delta y$ as given above is finite without renormalization.

We have verified that our result is gauge invariant by calculating $\delta y$ in $R_\xi$ gauge and expanding the result in powers of the Higgs mass $m_h$. The leading term $\delta y(m_h=0)$ in $R_\xi$ gauge corresponds to $\delta y$ in unitarity gauge. We have checked that the remainder, $\delta y(m_h)-\delta y(m_h=0)$, is numerically subleading in the parameter space where $\chi_b^0$ is the lightest neutral state and satisfies the relic density constraint (see Fig.~\ref{fig:majdt}). This indicates that electroweak box diagrams, which contribute to the effective scalar interaction $G_h^q$ in Eq.~\eqref{eq:dd-eff} at the same order of $m_h$ as this remainder, are suppressed with respect to the leading Higgs vertex contributions, which enter $\propto m_h^{-2}$ in the amplitude. Our calculation has been performed with two independent computer codes, one of which is based on the programs {\tt FeynArts}~\cite{Hahn:2000kx} and {\tt FeynCalc}~\cite{Mertig:1990an}. Our result agrees with Ref.~\cite{Dedes:2014hga}.

\bibliographystyle{apsrev}
\bibliography{fswz_ref}

\end{document}